\documentclass[12pt,onecolumn,english]{IEEEtran}
\usepackage[T1]{fontenc}
\usepackage{babel}
\usepackage{verbatim}
\usepackage{amsmath}
\usepackage{amsthm}
\usepackage{amssymb}
\usepackage{graphicx}
\PassOptionsToPackage{normalem}{ulem}
\usepackage{ulem}
\usepackage[unicode=true,
 bookmarks=true,bookmarksnumbered=true,bookmarksopen=true,bookmarksopenlevel=1,
 breaklinks=false,pdfborder={0 0 0},backref=false,colorlinks=false]
 {hyperref}
\hypersetup{
 pdfauthor={Kaniska Mohanty,  Mahesh Varanasi},
 pdfpagelayout=OneColumn, pdfnewwindow=true, pdfstartview=XYZ, plainpages=false}

\makeatletter
  \theoremstyle{plain}
  \newtheorem{lem}{\protect\lemmaname}
  \theoremstyle{plain}
  \newtheorem{thm}{\protect\theoremname}
  \theoremstyle{plain}
  \newtheorem{cor}{\protect\corollaryname}
  \theoremstyle{remark}
  \newtheorem{rem}{\protect\remarkname}

\addto\captionsenglish{}

 \DeclareMathOperator{\tr}{tr}

\@ifundefined{showcaptionsetup}{}{%
 \PassOptionsToPackage{caption=false}{subfig}}
\usepackage{subfig}
\makeatother

\providecommand{\corollaryname}{Corollary}
\providecommand{\lemmaname}{Lemma}
\providecommand{\remarkname}{Remark}
\providecommand{\theoremname}{Theorem}

\begin{document}

\title{The Generalized Degrees of Freedom Region of the MIMO Z-Interference
Channel with Delayed CSIT}

\author{Kaniska Mohanty and Mahesh K. Varanasi
~\thanks{This work was supported in part by NSF Grant 1423657. The authors are with the Department of Electrical, Computer, and Energy
Engineering, University of Colorado, Boulder, CO 80309-0425, e-mail:
\protect\href{http://kaniska.mohanty, varanasi@colorado.edu}{kaniska.mohanty, varanasi@colorado.edu}.}}
\maketitle
\begin{abstract}
The generalized degrees of freedom (GDoF) region of the multiple-input
multiple-output (MIMO) Gaussian Z-interference channel with an arbitrary
number of antennas at each node is established under the assumption
of delayed channel state information at transmitters (CSIT). The GDoF
region is parameterized by $\alpha$, which links the interference-to-noise
ratio (INR) to the signal-to-noise ratio (SNR) via $\mathrm{INR}=\mathrm{SNR}^{\alpha}$.
A new outer bound for the GDoF region is established by maximizing
a bound on the weighted sum-rate of the two users, which in turn is obtained
by using a combination of genie-aided side-information and an extremal
inequality. The maximum weighted sum-rate in the high SNR regime is shown to occur
when the transmission covariance matrix of the interfering transmitter
has full rank. An achievability scheme based on block-Markov encoding
and backward decoding is developed which uses interference quantization
and digital multicasting to take advantage of the channel statistics
of the cross-link, and the scheme is separately shown to be GDoF-optimal
in both the weak ($\alpha\leq1$) and strong $\left(\alpha>1\right)$
interference regimes. This is the first complete characterization
of the GDoF region of any interference network with
delayed CSIT, as well as the first such GDoF characterization of a
MIMO network with delayed CSIT and arbitrary number of antennas at
each node. For all antenna tuples, the GDoF region is shown to be
equal to or larger than the degrees of freedom (DoF) region over the
entire range of $\alpha$, which leads to a V-shaped
maximum sum-GDoF as a function of $\alpha$, with the minimum occurring
at $\alpha=1$. The delayed CSIT GDoF region and the sum-DoF are compared
with their counterparts under perfect CSIT, thereby
characterizing all antenna tuples and ranges of $\alpha$ for which
delayed CSIT is sufficient to achieve the perfect CSIT GDoF region
(or sum-DoF). It is also shown that treating interference as noise
is not, in general, GDoF-optimal for the MIMO Z-IC, even in the weak
interference regime.\end{abstract}


\newpage

\section{INTRODUCTION}

\IEEEPARstart{ T}{he} Z-interference channel (Z-IC) models two
transmitter/receiver pairs communicating over a shared medium, such
that only one transmitter causes interference at its unpaired receiver.
It is hence the simplest model that incorporates all of the
following salient features of wireless communications: broadcast,
superposition, distributed transmission and links with disparate strengths.
A study of the Z-IC is therefore tantamount to the study of the simplest setting 
in which these features are all simultaneously present in a wireless network.
From a GDoF perspective, the Z-IC can be used to model a two-user interference
channel (IC) in which the interference strength 
over one of the cross links is sufficiently weak compared to that of the other links 
that it can be mathematically modeled as the regime in which the high-SNR limit of the ratio of 
$\mathrm{INR}$ to $\mathrm{SNR}$, each measured in the dB scale, is zero.

The delayed CSIT model was introduced for the $K$-user multiple-input
single-output (MISO) broadcast channel (BC) in \cite{Maddah-Ali2012},
where it was shown that there is a significant DoF advantage (over the no CSIT
scenario) when the transmitter knows the previous channel states,
even when these states are independent of the unknown current channel
state. Delayed CSIT can arise in mobile communication scenarios with
short coherence blocks (caused by a rapidly varying mobile environment)
where CSIT through feedback links can be outdated. So far, research
on delayed CSIT has mainly focused on the \emph{DoF} region of various networks,
in which all relevant communication links are implicitly assumed to be statistically
comparable in strength. This includes the characterization of the DoF
regions of the two-user MIMO BC, Z-IC and IC with delayed
CSIT in \cite{Vaze2011}, \cite{Mohanty2015} and \cite{Vaze2012a}, respectively. 
Achievable DoF results (without converses) for SISO interference
and X networks with delayed CSIT were obtained in \cite{Abdoli2013},
and bounds on the DoF for the symmetric MIMO three-user BC were obtained
in \cite{Abdoli2011}. Under the constraint of linear encoding strategies
with delayed CSIT, the sum-DoF of the two-user MIMO X-channel
was characterized in 
\cite{Kao2014}, as was the DoF region of the two-user symmetric
MIMO X-channel. 

In practical wireless networks however, signal strengths are disparate in general.
In such settings, the \emph{G}DoF metric is eminently more suitable. Introduced in the context of the single-input single-output
(SISO) IC in \cite{Etkin2008}, GDoF measures the rate of linear growth of the
capacity region relative to a nominal $\log\:\mathrm{SNR}$ with increasing nominal SNR, 
when the SNRs and INRs are assigned different {\em exponents} with respect to the nominal SNR.
The GDoF region of the SISO IC was found in \cite{Etkin2008}, a result that was 
later generalized in \cite{Karmakar2012a}, to the MIMO IC with an arbitrary number
of antennas at each node. Both these works  assumed perfect
and instantaneous CSIT.

The first step towards studying GDoF under
channel uncertainty was taken in \cite{Vaze2011a}, wherein the GDoF region
of the MIMO IC with no CSIT
was characterized for the weak interference regime under certain antenna
configurations. Later, \cite{Karmakar2012} showed that, 
in the slow-fading MIMO IC, the generalized multiplexing gain region 
of the MIMO IC with perfect CSIT could be achieved by using {\em quantized} CSIT, 
with a sufficiently fast scaling of the number of feedback bits with the INR. 

More recently, \cite{Chen2015} studied the GDoF
region of the two-user MISO BC with fixed and alternating topologies
(link strengths) under alternating CSIT, where the CSIT state of each
link can be either perfect and instantaneous (P), delayed (D) or not
known (N). Inner and outer bounds on the GDoF under delayed CSIT were obtained therein, which 
are, however, not tight, so that the characterization of the GDoF region of 
even a two-user MISO BC with a static topology and delayed CSIT remains
an open problem.

The contributions of this paper can be summarized as follows.
The GDoF region of the MIMO Z-IC under the assumption
of delayed CSIT is established, providing
the first complete GDoF characterization of a {\em MIMO} network
under the delayed CSIT assumption (and with an arbitrary number of antennas at each node)
and also the first one obtained for an {\em interference} network (with distributed transmitters) 
and delayed CSIT. By using a genie-aided technique from \cite{Vaze2012a}
to provide side-information to one of the receivers and by applying the
extremal inequality from \cite{Liu2007}, we bound a weighted sum-rate
of the two users from above. We prove that, for both weak interference
$\left(\alpha\leq1\right)$ and strong interference $\left(\alpha>1\right)$, 
(with $\alpha=\frac{\log\:\mathrm{INR}}{\log\:\mathrm{SNR}}$), 
the asymptotic approximation of this bound in the high SNR regime
is maximized when the involved covariance matrix has full rank, thereby obtaining
a new outer bound for the GDoF region. We
also design a new achievability scheme with block-Markov encoding
and backward decoding that is able to achieve this outer bound in
both the weak and strong interference regimes. The achievability scheme
uses interference quantization in each block to compress the previous
block's interference, which in turn is reconstructed using delayed CSIT. 
It then digitally multicasts the quantization index of the previous interference as a common
message for both receivers.
Decoding starts from the final block, proceeding
successively backwards to the first block, using the common message
decoded in block $b+1$ as side-information while decoding block $b$.
By specifying the GDoF carried by the common and private messages,
as well as the transmit power at each transmitter, all the corner
points of the GDoF outer bound region in both the strong and weak
interference regimes are shown to be achievable. The GDoF region is
found to be equal to or larger than the corresponding DoF region $\left(\alpha=1\right)$
for all antenna tuples in both the weak and strong interference regimes,
thus demonstrating the advantage of incorporating knowledge of disparity
in link strengths under delayed CSIT. The sum-GDoF, as a function
of $\alpha$, is shown to be V-shaped, with the minimum occurring
at $\alpha=1$. Moreover, by comparing the delayed CSIT GDoF region
and sum-DoF with their perfect CSIT counterparts, we characterize
all the antenna-tuples and ranges of $\alpha$ where delayed CSIT
is sufficient to attain the perfect CSIT GDoF region and sum-GDoF.
We also illustrate, with an example, that treating interference as
noise is not always a GDoF-optimal strategy for the MIMO Z-IC with
delayed CSIT.

It is notable that the quantization and multicasting of the interference 
were also used in DoF-optimal schemes for the so-called mixed CSIT model, where
each transmitter has access to an imperfect estimate of the current
channel in addition to accurate delayed CSIT. This commonality 
is not surprising however,
since mixed CSIT achievability schemes, like GDoF schemes, must manage interference 
in the signal power level
dimension. In the case of mixed CSIT, these issues arise because transmit
beamforming in the null space of the (imperfect) current channel estimate
results in different power levels for the interference and the signal
at a receiver, a situation that also arises in the GDoF model because
of the inherent disparity in the strengths of different links. 

In particular, \cite{Yi2014} presents an achievability scheme under the mixed CSIT model with
features similar to ours, e.g., block-Markov encoding, introduced originally
in \cite{Cover1979} for the relay channel and later for channels
involving feedback in \cite{Ozarow1984} and \cite{Suh2011}, and backward decoding and
interference quantization, also used in earlier works on
mixed CSIT, e.g., \cite{Yang2012}, \cite{Gou2012} and \cite{Chen2012}.

The GDoF-optimal scheme presented in this work under delayed CSIT, however,
differs from the scheme in \cite{Yi2014} in several ways. Unlike
the mixed CSIT model, where the interference can only be attenuated
(through beamforming at the transmitters), the GDoF model allows for
interference that is stronger than the desired signal, when $\alpha>1$.
Similarly, the common message in each block, which carries the quantization
index of the compressed interference from the previous block, is always
received at a fixed power level in the achievable scheme from \cite{Yi2014},
but varies with $\alpha$ in the GDoF achievable scheme. Thus, it
is clear that incorporating the channel statistics of the interfering
link in a GDoF-optimal achievable scheme
requires a different analysis of the achievable GDoF region compared
to the mixed CSIT DoF region. Furthermore, the GDoF regions in the
weak and strong interference regimes have different corner points
and need to be considered separately, and require separate transmit
power and common message rate allocations to achieve those corner points. 
Moreover, absence of any current channel estimate
in the delayed CSIT model precludes any kind of transmitter beamforming,
a significant
feature of the mixed CSIT achievability schemes. Also, in a MIMO system,
transmit beamforming with mixed CSIT can attenuate the power level
of only those data streams that are in the null space of the channel
estimate of a cross-link (when such a null space exists), while in the
delayed CSIT GDoF model the power level of \emph{all} data
streams received over the cross-link are equally affected, and are even
strengthened when $\alpha>1$.
Other related work on mixed CSIT includes the characterization of
the DoF region of the MISO BC in \cite{Yang2012}, \cite{Gou2012}
and \cite{Chen2012}, and the MIMO Z-IC in \cite{Mohanty2015b}.

The rest of the paper is organized as follows. 
The channel model is described in the next section. The main result
of this paper is presented in Section \ref{sec:MAIN-RESULTS}, followed
by the proof of the outer bound in Section \ref{sec:PROOF-OF-OUTER-BOUND}.
In Section \ref{sec:ACHIEVABLE-SCHEME}, we describe the general achievability
scheme, which is shown to be GDoF optimal for both the weak interference
and strong interference regimes, in Sections \ref{sec:WEAK-INTERFERENCE}
and \ref{sec:STRONG-INTERFERENCE}, respectively. Various aspects
of the main result are discussed in Section \ref{sec:INSIGHT}, and
we conclude the paper in Section \ref{sec:CONCLUSION}.

The notation used in this paper is as follows: $\mathbb{R}$
and $\mathbb{Z}^{+}$ refer to the set of real numbers and non-negative
integers, respectively. Logarithm to the base 2 is denoted by $\log\left(\right)$.
The conjugate transpose of a matrix $A$ is denoted as $A^{\dagger}$,
its determinant as $\left|A\right|$ and the trace of $A$ as $\tr\left(A\right)$.
$\mathbf{I}_{n}$ is the identity matrix of size $n\times n$. For
two matrices $A$ and $B$, $A\preceq B$ means that the matrix $B-A$
is positive semi-definite. $\left(x\right)^{+}$ refers to the maximum
of a real number $x$ and 0. $\mathcal{C\mathcal{N}}\left(0,Q\right)$
refers to the distribution of complex circularly symmetric Gaussian
random vector with zero mean and covariance matrix $Q$. $\mathbb{E}\left[X\right]$
is the expectation of a random variable $X$. We use the standard
Landau notation, where $\mathcal{O}\left(1\right)$ refers to any
quantity that is bounded above by a constant. The approximation $g\left(\rho\right)\sim h\left(\rho\right)$
is a shorthand for $\lim_{\rho\rightarrow\infty}\frac{g\left(\rho\right)}{h\left(\rho\right)}=C,$
where $C$ is a constant that does not scale with $\rho$.

\section{THE CHANNEL MODEL \label{sec:THE-CHANNEL-MODEL}}

\begin{figure}[tb]
\includegraphics[scale=0.6]{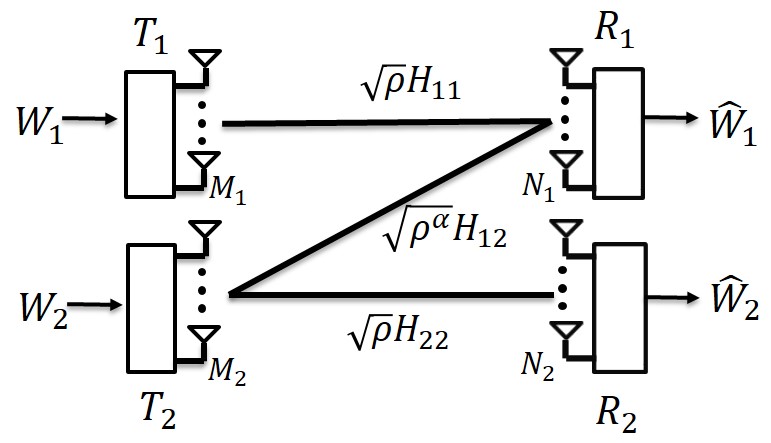}

\caption{\label{fig:ZIC-model}The $\left(M_{1},M_{2},N_{1},N_{2}\right)$
MIMO Z-IC.}

\end{figure}
The $\left(M_{1},M_{2},N_{1},N_{2}\right)$ MIMO Z-IC consists of
two transmitters $T_{1}$ and $T_{2}$ with $M_{1}$ and $M_{2}$
antennas, respectively, and their paired receivers $R_{1}$ and $R_{2}$,
with $N_{1}$ and $N_{2}$ antennas, respectively. Each transmitter
$T_{i}$ sends a unicast message $W_{i}$ to its paired receiver $R_{i}$,
$i\in\left\{ 1,2\right\} $. $T_{2}$ causes interference at $R_{1}$,
but $T_{1}$ does not cause any interference at $R_{2}$ (see Fig.
\ref{fig:ZIC-model}). The received signals at the two receivers at
time $t$ are as follows: 
\begin{eqnarray*}
Y_{1}\left(t\right) & = & \sqrt{\rho}H_{11}\left(t\right)X_{1}\left(t\right)+\sqrt{\rho^{\alpha}}H_{12}\left(t\right)X_{2}\left(t\right)+Z_{1}\left(t\right),\\
Y_{2}\left(t\right) & = & \sqrt{\rho}H_{22}\left(t\right)X_{2}\left(t\right)+Z_{2}\left(t\right),
\end{eqnarray*}
where $X_{i}\left(t\right)\in\mathbb{C}^{M_{i}\times1}$ is the transmitted
signal from $T_{i}$, $Y_{i}\left(t\right)\in\mathbb{C}^{N_{i}\times1}$
is the received signal at $R_{i}$, $H_{ji}\left(t\right)\in\mathbb{C}^{N_{j}\times M_{i}}$
is the channel matrix from $T_{i}$ to $R_{j}$, $Z_{i}\left(t\right)\sim\mathcal{CN}\left(0,\mathbf{I}_{N_{i}}\right)$
is the additive Gaussian noise (with unit variance) at $R_{i}$, and
$\rho$ and $\rho^{\alpha}$, where $\rho>0$ and $\alpha\geq0$,
are the channel gains of the direct links and interfering link, respectively,
for $i\in\left\{ 1,2\right\} $. Transmitter $T_{i}$ has an average
power constraint $\tr\left(Q_{i}\right)\leq1$, where $Q_{i}\triangleq\mathbb{E}\left(X_{i}X_{i}^{\dagger}\right)$,
$i\in\left\{ 1,2\right\} $. All entries of all channel matrices are
independent and identically distributed (i.i.d.), and the channel
matrices and noise are assumed to be i.i.d. complex Gaussian (with
unit variance) across time and independent of each other. Thus, the
INR at $R_{1}$ is $\rho^{\alpha}$ while the SNR at both receivers
is $\rho$. We define the interference to be weak when $\alpha\leq1$,
and the interference is said to be strong when $\alpha>1$. We also
define $\mathcal{H}\left(t\right)\triangleq\left\{ H_{11}\left(t\right),H_{12}\left(t\right),H_{22}\left(t\right)\right\} $,
and all channel matrices up to time $\tau$ are denoted by $\mathcal{H}^{\tau}\triangleq\left\{ \mathcal{H}\left(t\right)\right\} _{t=1}^{\tau}$. 

Both receivers have perfect knowledge of all the channel matrices.
In other words, the decoding function at receiver $R_{i}$, $\forall i\in\left\{ 1,2\right\} $,
for a codeword spanning $n$ channel uses is $g_{i}\left(\left\{ Y_{i}\left(t\right)\right\} _{t=1}^{n},\mathcal{H}^{n}\right)=\hat{W}_{i}$,
where $\hat{W}_{i}$ is the decoded message at $R_{i}$. The transmitters,
on the other hand, learn the channel matrices only after a finite
delay which, without loss of generality, we assume to be 1. Thus,
at time $t$, each transmitter knows all the channel matrices up to
time $t-1$. This is known as the delayed CSIT assumption. Consequently,
the encoding function for $T_{i}$, $i\in\left\{ 1,2\right\} $, at
time $t$ is $h_{i,t}\left(W_{i},\mathcal{H}^{t-1}\right)$. 

The rate tuple $\left(\bar{R}_{1}\left(\rho,\alpha\right),\bar{R}_{2}\left(\rho,\alpha\right)\right)$,
where $\bar{R}_{i}=\frac{\log\left|\mathcal{W}_{i}\right|}{n}$ and
$\left|\mathcal{W}_{i}\right|$ is the cardinality of the message
set $\mathcal{W}_{i}$ at $T_{i}$, is said to be achievable if there
exists a codeword spanning $n$ channel uses such that the probability
of error goes to zero as $n\rightarrow\infty$. The capacity region
$C\left(\rho,\alpha\right)$ is the region of all such achievable
rate tuples, and the GDoF region is defined as the pre-log factor
of the capacity region as $\rho\rightarrow\infty$, i.e., 
\[
\mathbf{D}=\biggl\{(d_{1},d_{2})\biggl|\ d_{i}\geq0\ {\rm and}\ \exists\ \left(\bar{R}_{1}(\rho,\alpha),\bar{R}_{2}(\rho,\alpha)\right)\in C(\rho,\alpha),
\]
\[
\left.\text{\text{such that}}\ d_{i}=\lim_{\rho\rightarrow\infty}\frac{\bar{R}_{i}(\rho,\alpha)}{\log(\rho)}\:,i\in\{1,2\}\right\} .
\]
We also define the sum-GDoF as follows:
\[
d_{\sum}=\sup\left\{ d_{1}+d_{2}\left|\left(d_{1},d_{2}\right)\in\mathbf{D}\right.\right\} .
\]

Note that in our notation $R_j$ denotes the $j^{\rm th}$ receiver and $ \bar{R}_{j}$ denotes the $j^{\rm th}$ user's information 
rate.

\section{MAIN RESULT\label{sec:MAIN-RESULTS}}

In the following lemma, borrowed from \cite{Karmakar2012a}, we
define a function $f\left(\right)$ which provides an approximation
(up to an $\mathcal{O}\left(1\right)$ term) of the sum-rate upper
bound of the 2-user MIMO multiple-access channel (MAC) in the asymptotically
high SNR regime. The function $f()$ will be useful not only in stating
the GDoF region of the MIMO Z-IC in Theorem \ref{thm:Main-Result},
but also in obtaining asymptotic approximations throughout the paper.
\begin{lem}
\label{lem:MAC-approx}Let $H_{1}\in\mathbb{C}^{u\times u_{1}}$ and
$H_{2}\in\mathbb{C}^{u\times u_{2}}$ be two full rank (with probability
1) channel matrices such that the matrix $H\triangleq\left[H_{1}\;H_{2}\right]$
is also full rank (with probability 1). Then, for $\rho\rightarrow\infty$,
we have
\begin{eqnarray*}
 & \log\left|\mathbf{I}_{u}+\rho^{a_{1}}H_{1}H_{1}^{\dagger}+\rho^{a_{2}}H_{2}H_{2}^{\dagger}\right|\\
= & f\left(u,\left(a_{1},u_{1}\right),\left(a_{2},u_{2}\right)\right)\log\left(\rho\right)+\mathcal{O}\left(1\right),
\end{eqnarray*}
where for any $\left(u,u_{1},u_{2}\right)\in\mathbb{Z}^{+3}$ and
$\left(a_{1},a_{2}\right)\in\mathbb{R}^{2}$, the function $f$ is
defined as

\begin{eqnarray*}
f\left(u,\left(a_{1},u_{1}\right),\left(a_{2},u_{2}\right)\right) & = & \min\left(u,u_{i_{1}}\right)a_{i_{1}}^{+}\\
 &  & +\min\left(\left(u-u_{i_{1}}\right)^{+},u_{i_{2}}\right)a_{i_{2}}^{+},
\end{eqnarray*}
for $i_{1}\neq i_{2}\in\left\{ 1,2\right\} $ such that $a_{i_{1}}\geq a_{i_{2}}$.
\end{lem}
The next theorem states the main result of this paper.
\begin{thm}
\label{thm:Main-Result}The GDoF region of the MIMO Z-IC with delayed
CSIT is given by the following set of inequalities:
\begin{eqnarray}
 & d_{1}\leq\min\left(M_{1},N_{1}\right),\quad d_{2}\leq\min\left(M_{2},N_{2}\right),\label{eq:PTP-bound}\\
 & \frac{d_{1}}{\min\left(M_{2},N_{1}\right)}+\frac{d_{2}}{\min\left(M_{2},N_{1}+N_{2}\right)}\leq\frac{f\left(N_{1},\left(\alpha,M_{2}\right),\left(1,M_{1}\right)\right)}{\min\left(M_{2},N_{1}\right)}+\nonumber \\
 & \qquad\qquad\qquad\qquad\qquad+\frac{f\left(M_{2},\left(\alpha,N_{1}\right),\left(1,N_{2}\right)\right)}{\min\left(M_{2},N_{1}+N_{2}\right)}-\alpha.\label{eq:delayed-CSIT-bound}
\end{eqnarray}
\end{thm}
\begin{IEEEproof}
The inequalities in \eqref{eq:PTP-bound} are the single-user outer
bounds for the individual MIMO point-to-point channels. The outer
bound \eqref{eq:delayed-CSIT-bound} is proved in the next section.%

The general achievability scheme is described in Section \ref{sec:ACHIEVABLE-SCHEME},
and it is shown to achieve the GDoF outer bound region in both the
weak and strong interference regimes, in Sections \ref{sec:WEAK-INTERFERENCE}
and \ref{sec:STRONG-INTERFERENCE}, respectively.
\end{IEEEproof}
When $M_{2}>N_{1}+N_{2}$, it is clear that the GDoF region in Theorem \ref{thm:Main-Result}
does not depend on $M_{2}$. In this case,
without loss of generality,
we can switch off the extra transmit antennas at $T_{2}$ and prove
achievability of the GDoF region using the general achievability scheme
from Section \ref{sec:ACHIEVABLE-SCHEME} with $M_{2}=N_{1}+N_{2}$.
Thus, from an achievability point of view, we can always assume $M_{2}\leq N_{1}+N_{2}$.
Using similar arguments, the general achievability scheme in Section
\ref{sec:ACHIEVABLE-SCHEME} can be restricted (without loss of generality)
to the following antenna configurations:
\begin{eqnarray}
M_{1}\leq N_{1}, &  & N_{1}\leq M_{1}+M_{2},\nonumber \\
N_{2}\leq M_{2}, &  & M_{2}\leq N_{1}+N_{2}.\label{eq:gdof-zic-antenna-assumptions-end}
\end{eqnarray}
Define $N_{1}^{\prime}\triangleq\min\left(M_{2},N_{1}\right)$.
The following corollary specifies the sum-GDoF.
\begin{cor}
\label{cor:sum-GDoF}Under the antenna assumptions \eqref{eq:gdof-zic-antenna-assumptions-end},
the sum-GDoF of the MIMO Z-IC with delayed CSIT when $\alpha\leq1$
is as follows: 
\begin{equation}
\negthickspace\negthickspace d_{\sum}=\min\negmedspace\left(\negmedspace M_{1}\negmedspace+\negmedspace N_{2},\medspace M_{1}\negmedspace+\negmedspace N_{2}\negmedspace-\negmedspace\alpha\left(M_{1}\negmedspace-\negmedspace N_{1}\negmedspace+\negmedspace\frac{N_{2}N_{1}^{\prime}}{M_{2}}\right)\right)\negmedspace,\label{eq:sum-GDoF-weak-int}
\end{equation}
and the sum-GDoF when $\alpha>1$ is as follows: 
\begin{equation}
\negmedspace\negmedspace d_{\sum}=\min\negmedspace\left(\negmedspace M_{1}\negmedspace+\negmedspace N_{2},\medspace N_{2}\negmedspace+\negmedspace N_{1}\negmedspace-\negmedspace\frac{\left(N_{2}+N_{1}^{\prime}\right)N_{1}^{\prime}}{M_{2}}\negmedspace+\negmedspace\frac{\left(N_{1}^{\prime}\right)^{2}}{M_{2}}\alpha\right)\negmedspace.\label{eq:sum-GDoF-strong-int}
\end{equation}
\end{cor}
\begin{IEEEproof}
In Sections \ref{sec:WEAK-INTERFERENCE} and \ref{sec:STRONG-INTERFERENCE},
we show that \eqref{eq:sum-GDoF-weak-int} and \eqref{eq:sum-GDoF-strong-int},
respectively, are direct consequences of Theorem \ref{thm:Main-Result}
and the antenna assumptions in \eqref{eq:gdof-zic-antenna-assumptions-end},
when $\alpha\leq1$ and $\alpha>1$, respectively.
\end{IEEEproof}

\section{PROOF OF OUTER BOUND \label{sec:PROOF-OF-OUTER-BOUND}}

To prove outer bound \eqref{eq:delayed-CSIT-bound}, we begin by using
the genie-aided technique from \cite{Vaze2012a} and \cite{Vaze2012},
where side-information about $R_{1}$'s message and received signal
is provided to $R_{2}$. A physically degraded channel is thus obtained, and Fano's inequality is used to bound the individual rate of each user. A weighted sum of these individual rates is, in turn, bounded using the extremal inequality from \cite{Liu2007} for physically degraded channels, by following the steps in \cite{Yi2014} and \cite{Yang2012}. Subsequently,we show that the asymptotic approximation of this weighted sum-rate
bound at high SNR, obtained by applying Lemma \ref{lem:MAC-approx},
is maximized, for all antenna configurations and values of $\alpha$,
when the transmit covariance matrix at $T_{2}$ is full rank. This
maximization yields the delayed CSIT GDoF outer bound \eqref{eq:delayed-CSIT-bound}.

We define the following virtual received signals,
obtained by subtracting the effect of $X_{1}$ from the received signal
at each receiver:
\begin{eqnarray*}
\bar{Y}_{1}\left(t\right) & \triangleq & \sqrt{\rho^{\alpha}}H_{12}\left(t\right)X_{2}\left(t\right)+Z_{1}\left(t\right),\\
\bar{Y}_{2}\left(t\right) & \triangleq & \sqrt{\rho}H_{22}\left(t\right)X_{2}\left(t\right)+Z_{2}\left(t\right),
\end{eqnarray*}
and introduce the notation $\bar{Y}_{i}^{k}\triangleq\left\{ \bar{Y}_{i}\left(t\right)\right\} _{t=1}^{k}$,
$Y_{i}^{k}\triangleq\left\{ Y_{i}\left(t\right)\right\} _{t=1}^{k}$
and $X_{i}^{k}\triangleq\left\{ X_{i}\left(t\right)\right\} _{t=1}^{k}$,
$\forall i\in\left\{ 1,2\right\} $. Note that $Y_{2}\left(t\right)$
and $\bar{Y}_{2}\left(t\right)$ are exactly the same, but we use
the above notation anyway, for convenience. Since the probability
of error $P_{e}^{\left(n\right)}$ goes to zero as $n\rightarrow\infty$,
we denote $n\epsilon_{n}\triangleq1+n\bar{R}P_{e}^{\left(n\right)}$
so that $\lim_{n\rightarrow\infty}\epsilon_{n}=0$.

We next bound the achievable rate for the first user as follows:
\begin{eqnarray}
 &  & n\left(\bar{R}_{1}-\epsilon_{n}\right)\nonumber \\
 &  & \overset{\left(a\right)}{\leq}I\left(W_{1};Y_{1}^{n}\left|\mathcal{H}^{n}\right.\right)\nonumber \\
 &  & =I\left(W_{1},W_{2};Y_{1}^{n}\left|\mathcal{H}^{n}\right.\right)-I\left(W_{2};Y_{1}^{n}\left|W_{1},\mathcal{H}^{n}\right.\right)\nonumber \\
 &  & \overset{\left(b\right)}{=}h\left(Y_{1}^{n}\left|\mathcal{H}^{n}\right.\right)-I\left(W_{2};Y_{1}^{n}\left|W_{1},\mathcal{H}^{n}\right.\right)+n\mathcal{O}\left(1\right)\nonumber \\
 &  & \overset{\left(c\right)}{\leq}\sum_{t=1}^{n}h\left(Y_{1}\left(t\right)\left|\mathcal{H}\left(t\right)\right.\right)-h\left(Y_{1}^{n}\left|W_{1},\mathcal{H}^{n}\right.\right)+n\mathcal{O}\left(1\right)\nonumber \\
 &  & \overset{\left(d\right)}{=}\sum_{t=1}^{n}h\left(Y_{1}\left(t\right)\left|\mathcal{H}\left(t\right)\right.\right)-h\left(\bar{Y}_{1}^{n}\left|\mathcal{H}^{n}\right.\right)+n\mathcal{O}\left(1\right)\nonumber \\
 &  & \overset{\left(e\right)}{\leq}\sum_{t=1}^{n}h\left(Y_{1}\left(t\right)\left|\mathcal{H}\left(t\right)\right.\right)\nonumber \\
 &  & \quad-\sum_{t=1}^{n}h\left(\bar{Y}_{1}\left(t\right)\left|\mathcal{H}^{n},\bar{Y}_{1}^{t-1},\bar{Y}_{2}^{t-1}\right.\right)+n\mathcal{O}\left(1\right)\nonumber \\
 &  & \overset{\left(f\right)}{=}\sum_{t=1}^{n}h\left(Y_{1}\left(t\right)\left|\mathcal{H}\left(t\right)\right.\right)\nonumber \\
 &  & \quad-\sum_{t=1}^{n}h\left(\bar{Y}_{1}\left(t\right)\left|\mathcal{U}\left(t\right),\mathcal{H}\left(t\right)\right.\right)+n\mathcal{O}\left(1\right),\label{eq:outer-bound-proof-d1}
\end{eqnarray}
where $\left(a\right)$ follows from Fano's inequality, $\left(b\right)$
and $\left(c\right)$ are true because $Y_{1}^{n}$ is a deterministic function
(up to $\mathcal{O}\left(1\right)$ approximation) of $W_{1}$, $W_{2}$
and $\mathcal{H}^{n}$. We obtain $\left(d\right)$
by removing the effect of $W_{1}$ from $Y_{1}^{n}$ to obtain the
virtual signal $\bar{Y}_{1}^{n}$, and $\left(e\right)$ uses the
fact that conditioning decreases entropy. Finally, we define $\mathcal{U}\left(t\right)\triangleq\left\{ \bar{Y}_{1}^{t-1},\bar{Y}_{2}^{t-1},\mathcal{H}^{t-1}\right\} $
and the equality $\left(f\right)$ holds because, given $\mathcal{U}\left(t\right)$
and $\mathcal{H}\left(t\right)$, $\bar{Y}_{1}\left(t\right)$ is
independent of $\left\{ \mathcal{H}\left(\tau\right)\right\} _{\tau=t+1}^{n}$.

Next, as seen in \cite{Vaze2012a} and \cite{Vaze2012}, a genie provides receiver $R_{2}$ with both $R_{1}$'s message
$W_{1}$ and the received signals $Y_{1}^{n}$ at $R_{1}$. Now, using
Fano's inequality again, the achievable rate for user 2 is bounded as follows:
\begin{eqnarray}
 &  & n\left(\bar{R}_{2}-\epsilon_{n}\right)\nonumber \\
 &  & \overset{}{\leq}I\left(W_{2};Y_{1}^{n},Y_{2}^{n},W_{1}\left|\mathcal{H}^{n}\right.\right)\nonumber \\
 &  & \overset{\left(a\right)}{=}I\left(W_{2};Y_{1}^{n},Y_{2}^{n}\left|W_{1},\mathcal{H}^{n}\right.\right)\nonumber \\
 &  & \overset{\left(b\right)}{=}I\left(W_{2};\bar{Y}_{1}^{n},\bar{Y}_{2}^{n}\left|\mathcal{H}^{n}\right.\right)\nonumber \\
 &  & \overset{}{=}\sum_{t=1}^{n}I\left(W_{2};\bar{Y}_{1}\left(t\right),\bar{Y}_{2}\left(t\right)\left|\bar{Y}_{1}^{t-1},\bar{Y}_{2}^{t-1},\mathcal{H}^{n}\right.\right)\nonumber \\
 &  & \overset{\left(c\right)}{\leq}\sum_{t=1}^{n}I\left(X_{2}\left(t\right);\bar{Y}_{1}\left(t\right),\bar{Y}_{2}\left(t\right)\left|\bar{Y}_{1}^{t-1},\bar{Y}_{2}^{t-1},\mathcal{H}^{n}\right.\right)\nonumber \\
 &  & =\sum_{t=1}^{n}\left(h\left(\bar{Y}_{1}\left(t\right),\bar{Y}_{2}\left(t\right)\left|\bar{Y}_{1}^{t-1},\bar{Y}_{2}^{t-1},\mathcal{H}^{n}\right.\right)\right.\nonumber \\
 &  & \qquad\left.-h\left(\bar{Y}_{1}\left(t\right),\bar{Y}_{2}\left(t\right)\left|X_{2}\left(t\right),\bar{Y}_{1}^{t-1},\bar{Y}_{2}^{t-1},\mathcal{H}^{n}\right.\right)\right)\nonumber \\
 &  & \overset{}{\leq}\sum_{t=1}^{n}h\left(\bar{Y}_{1}\left(t\right),\bar{Y}_{2}\left(t\right)\left|\bar{Y}_{1}^{t-1},\bar{Y}_{2}^{t-1},\mathcal{H}^{n}\right.\right)\nonumber \\
 &  & \overset{\left(d\right)}{=}\sum_{t=1}^{n}h\left(\bar{Y}_{1}\left(t\right),\bar{Y}_{2}\left(t\right)\left|\mathcal{U}\left(t\right),\mathcal{H}\left(t\right)\right.\right),\label{eq:outer-bound-proof-d2}
\end{eqnarray}
where $\left(a\right)$ is true because $W_{1}$ and $W_{2}$ are
independent of each other, $\left(b\right)$ is obtained by removing
the effect of $W_{1}$ from the received signals $Y_{1}^{n}$ and
$Y_{2}^{n}$, to obtain the virtual received signals $\bar{Y}_{1}^{n}$
and $\bar{Y}_{2}^{n}$, respectively, and the inequality in $\left(c\right)$
is an application of the data processing inequality. Finally, $\left(d\right)$
is true because, given $\mathcal{U}\left(t\right)$ and $\mathcal{H}\left(t\right)$,
both $\bar{Y}_{1}\left(t\right)$ and $\bar{Y}_{2}\left(t\right)$
are independent of $\left\{ \mathcal{H}\left(\tau\right)\right\} _{\tau=t+1}^{n}$.

We now define the following: 
\begin{eqnarray}
S\left(t\right) & \triangleq & \left[\begin{array}{c}
\sqrt{\rho^{\alpha}}H_{12}\left(t\right)\\
\sqrt{\rho}H_{22}\left(t\right)
\end{array}\right],\nonumber \\
K\left(t\right) & \triangleq & \mathbb{E}\left(x_{2}\left(t\right)x_{2}^{\dagger}\left(t\right)\left|\:\mathcal{U}\left(t\right)\right.\right),\nonumber \\
L\left(t\right) & \triangleq & \mathbb{E}\left(x_{1}\left(t\right)x_{1}^{\dagger}\left(t\right)\left|\:\mathcal{H}^{t-1}\right.\right),\nonumber \\
\mathcal{V}\left(t\right) & \triangleq & \left\{ \mathcal{U}\left(t\right),\mathcal{H}\left(t\right)\right\} ,\nonumber \\
p & \triangleq & \min\left(M_{2},N_{1}+N_{2}\right),\nonumber \\
q & \triangleq & \min\left(M_{2},N_{1}\right)\triangleq N_{1}^{\prime}.\label{eq:outer-bound-definitions}
\end{eqnarray}
 As shown in \cite{Yi2014} and \cite{Yang2012}, by applying the
extremal inequality for degraded outputs (see \cite{Liu2007}, \cite{Weingarten2006}) to the physically degraded channel $X_{2}^{n}\rightarrow\left(\bar{Y}_{1}^{n},\bar{Y}_{2}^{n}\right)\rightarrow\bar{Y}_{1}^{n}$,
we obtain the following inequality: 
\begin{eqnarray}
 &  & \frac{1}{p}h\left(\bar{Y}_{1}\left(t\right),\bar{Y}_{2}\left(t\right)\left|\mathcal{V}\left(t\right)\right.\right)-\frac{1}{q}h\left(\bar{Y}_{1}\left(t\right)\left|\mathcal{V}\left(t\right)\right.\right)\nonumber \\
 &  & \leq\negthickspace\negthickspace\negthickspace\max_{\begin{array}{c}
K\succeq0\\
\tr\left(K\right)\leq1
\end{array}}\negthickspace\negthickspace\mathbb{E}_{S}\left(\frac{1}{p}\log\left|\mathbf{I}_{N_{1}+N_{2}}+S\left(t\right)K\left(t\right)S^{\dagger}\left(t\right)\right|\right.\nonumber \\
 &  & \qquad\qquad\left.-\frac{1}{q}\log\left|\mathbf{I}_{N_{1}}+\rho^{\alpha}H_{12}\left(t\right)K\left(t\right)H_{12}^{\dagger}\left(t\right)\right|\right),\label{eq:outer-bound-proof-logdet}
\end{eqnarray}
which allows us to outer bound the weighted sum of the achievable
rate of the two users from \eqref{eq:outer-bound-proof-d1} and \eqref{eq:outer-bound-proof-d2}.
But first, we use Lemma \ref{lem:MAC-approx} to obtain the asymptotic
approximation in the high SNR regime of each of the terms in \eqref{eq:outer-bound-proof-logdet}.

Note that, because
of delayed CSIT, the covariance matrix $K\negthinspace\left(t\right)$
is independent of both $H_{12}\left(t\right)$ and $H_{22}\left(t\right)$,
and is thus also independent of $S\left(t\right)$. Now, let the matrix
$K\negthinspace\left(t\right)$ have rank $M_{2}-r$, such that $0\leq r\leq M_{2}$
with $r=0$ corresponding to a full rank $K\negmedspace\left(t\right)$. The singular
value decomposition (SVD) of the transmit covariance matrix $K\left(t\right)$
can be written as $K\left(t\right)=U\left(t\right)\Lambda\left(t\right)U^{\dagger}\left(t\right)$,
where $U\left(t\right)\in\mathbb{C}^{M_{2}\times\left(M_{2}-r\right)}$
is such that $U^{\dagger}U=\mathbf{I}$ and $\Lambda\left(t\right)$
is a $\left(M_{2}-r\right)\times\left(M_{2}-r\right)$ diagonal matrix
containing the non-zero singular values of $K\left(t\right)$ in descending
order. Using the SVD of $K\negmedspace\left(t\right)$, we obtain
the following asymptotic approximation of the first term in \eqref{eq:outer-bound-proof-logdet}
(we suppress the index $t$ henceforth): 
\begin{eqnarray}
 &  & \;\;\:\log\left|\mathbf{I}_{N_{1}+N_{2}}+SKS^{\dagger}\right|\nonumber \\
 &  & \overset{\left(a\right)}{=}\log\left|\mathbf{I}_{N_{1}+N_{2}}\negthickspace+\negthickspace\left[\begin{array}{c}
\sqrt{\rho^{\alpha}}\tilde{H}_{12}\\
\sqrt{\rho}\tilde{H}_{22}
\end{array}\right]\left[\begin{array}{c}
\sqrt{\rho^{\alpha}}\tilde{H}_{12}\\
\sqrt{\rho}\tilde{H}_{22}
\end{array}\right]^{\dagger}\right|\nonumber \\
 &  & \overset{\left(b\right)}{=}\log\left|\mathbf{I}_{M_{2}-r}+\rho^{\alpha}\tilde{H}_{12}^{\dagger}\tilde{H}_{12}+\rho\tilde{H}_{22}^{\dagger}\left(t\right)\tilde{H}_{22}\right|\nonumber \\
 &  & =f\left(M_{2}-r,\left(\alpha,N_{1}\right),\left(1,N_{2}\right)\right)\log\rho+n\mathcal{O}\left(1\right),\label{eq:outer-bound-proof-approx-1}
\end{eqnarray}
where, in $\left(a\right)$, we have defined $\tilde{H}_{i2}\negmedspace\triangleq\negmedspace H_{i2}U\Lambda^{\frac{1}{2}}$,
for $i\in\left\{ 1,2\right\} $, and in $\left(b\right)$, we use
the identity $\left|\mathbf{I}+AB\right|=\left|\mathbf{I}+BA\right|$
and finally, \eqref{eq:outer-bound-proof-approx-1} follows from Lemma
\ref{lem:MAC-approx}. 

Similarly, using the SVD of $K\left(t\right)$ again, we obtain the
following asymptotic approximation of the second term in \eqref{eq:outer-bound-proof-logdet}:%
\begin{align}
 & \log\left|\mathbf{I}_{N_{1}}\negmedspace+\negmedspace\rho^{\alpha}H_{12}KH_{12}^{\dagger}\right|\nonumber \\
 & =\alpha\min\left(M_{2}\negmedspace-\negmedspace r,N_{1}\right)\log\rho+n\mathcal{O}\left(1\right).\label{eq:outer-bound-proof-approx-2}
\end{align}

Next, we approximate the first term in \eqref{eq:outer-bound-proof-d1}
as follows: 
\begin{eqnarray}
 &  & \;\;\:\sum_{t=1}^{n}h\left(Y_{1}\left(t\right)\left|\mathcal{H}\left(t\right)\right.\right)\nonumber \\
 &  & \overset{\left(a\right)}{\leq}n\log\left|\mathbf{I}_{N_{1}}+\rho H_{11}LH_{11}^{\dagger}+\rho^{\alpha}H_{12}KH_{12}^{\dagger}\right|\nonumber \\
 &  & \overset{\left(b\right)}{\leq}n\log\left|\mathbf{I}_{N_{1}}+\rho H_{11}H_{11}^{\dagger}+\rho^{\alpha}\tilde{H}_{12}\tilde{H}_{12}^{\dagger}\right|\nonumber \\
 &  & \overset{\left(c\right)}{=}nf\negmedspace\left(N_{1},\left(1,M_{1}\right),\left(\alpha,M_{2}\negmedspace-\negmedspace r\right)\right)\log\rho\negthinspace+\negthinspace n\mathcal{O}\left(1\right)\negmedspace,\label{eq:outer-bound-proof-approx-3}
\end{eqnarray}
where $\left(a\right)$ uses the fact that Gaussian inputs maximize
the entropy. Since $\tr\left(L\right)\negmedspace\leq\negmedspace1$,
we have $L\negmedspace\preceq\mathbf{I}_{M_{1}}$ and $\left(b\right)$
follows by using the SVD of $K$ and substituting $L$ with $\mathbf{I}_{M_{1}}$,
since $\log\det$ is monotonically increasing on the cone of positive
definite matrices. The equality in $\left(c\right)$ is a direct consequence
of Lemma \ref{lem:MAC-approx}.

Finally, we outer bound the weighted sum of the achievable rates from
\eqref{eq:outer-bound-proof-d1} and \eqref{eq:outer-bound-proof-d2}
as follows:
\begin{eqnarray}
 &  & n\left(\frac{\bar{R}_{1}}{q}+\frac{\bar{R}_{2}}{p}-\epsilon_{n}\right)\nonumber \\
 &  & \overset{\left(a\right)}{\leq}n.\frac{1}{q}f\left(N_{1},\left(1,M_{1}\right),\left(\alpha,M_{2}-r\right)\right)\log\rho+n\mathcal{O}\left(1\right)\nonumber \\
 &  & \;+\frac{1}{p}\sum_{t=1}^{n}h\left(\bar{Y}_{1}\left(t\right),\bar{Y}_{2}\left(t\right)\left|\mathcal{V}\left(t\right)\right.\right)-\frac{1}{q}\sum_{t=1}^{n}h\left(\bar{Y}_{1}\left(t\right)\left|\mathcal{V}\left(t\right)\right.\right)\nonumber \\
 &  & \overset{\left(b\right)}{\leq}\frac{n}{q}f\left(N_{1},\left(1,M_{1}\right),\left(\alpha,M_{2}-r\right)\right)\log\rho\nonumber \\
 &  & \;+n\frac{f\left(M_{2}-r,\left(\alpha,N_{1}\right),\left(1,N_{2}\right)\right)}{p}\log\rho\nonumber \\
 &  & \;-n\frac{\alpha\min\left(M_{2}-r,N_{1}\right)}{q}\log\rho+n\mathcal{O}\left(1\right)\label{eq:outer-bound-proof-final-r-form}\\
 &  & \overset{\left(c\right)}{\leq}\frac{n}{q}f\left(N_{1},\left(1,M_{1}\right),\left(\alpha,M_{2}\right)\right)\log\rho\nonumber \\
 &  & +n\left(\negmedspace\frac{f\left(M_{2},\left(\alpha,N_{1}\right),\left(1,N_{2}\right)\right)}{p}-\frac{\alpha N_{1}^{\prime}}{q}\right)\negmedspace\log\rho+n\mathcal{O}\negmedspace\left(1\right),\nonumber \\
\label{eq:outer-bound-proof-final-f-form}
\end{eqnarray}
where inequality $\left(a\right)$ is obtained by using \eqref{eq:outer-bound-proof-d2}
and substituting \eqref{eq:outer-bound-proof-approx-3} in \eqref{eq:outer-bound-proof-d1},
and the second inequality $\left(b\right)$ is obtained by using \eqref{eq:outer-bound-proof-logdet},
after substituting the asymptotic approximations from \eqref{eq:outer-bound-proof-approx-1}
and \eqref{eq:outer-bound-proof-approx-2}. As shown below in Lemma
\ref{lem:rank-optimization-lemma}, the expression in \eqref{eq:outer-bound-proof-final-r-form}
is maximized for both weak and strong interference when $r=0$, and
thus, we substitute $r=0$ to obtain the inequality in $\left(c\right)$. Finally, substituting
the values of $p$ and $q$ from \eqref{eq:outer-bound-definitions}
in \eqref{eq:outer-bound-proof-final-f-form} and dividing both sides
of the inequality by $n\cdot\log\rho$ as $\rho\rightarrow\infty$
(for which $\epsilon_{n}\rightarrow0$), we obtain the outer bound
\eqref{eq:delayed-CSIT-bound}.
\begin{lem}
\label{lem:rank-optimization-lemma}The function $g\left(r\right)$,
defined as follows, 
\begin{eqnarray}
g\left(r\right) & \triangleq & \frac{f\left(N_{1},\left(1,M_{1}\right),\left(\alpha,M_{2}-r\right)\right)}{\min\left(M_{2},N_{1}\right)}\nonumber \\
 &  & +\:\frac{f\left(M_{2}-r,\left(\alpha,N_{1}\right),\left(1,N_{2}\right)\right)}{\min\left(M_{2},N_{1}+N_{2}\right)}\nonumber \\
 &  & -\:\alpha\frac{\min\left(M_{2}-r,N_{1}\right)}{\min\left(M_{2},N_{1}\right)},\label{eq:rank-opt-expression-to-optimize}
\end{eqnarray}
for $0\leq r\leq M_{2}$ and $r\in\mathbb{Z}$, is maximized at $r=0$,
for $\alpha\in\left[0,\infty\right)$.\end{lem}
\begin{IEEEproof}
See Appendix \ref{sec:Outer-Bound-Optimization}.
\end{IEEEproof}

\section{GENERAL ACHIEVABILITY SCHEME\label{sec:ACHIEVABLE-SCHEME}}

As explained in Section \ref{sec:MAIN-RESULTS}, we can restrict
the achievability scheme, without any loss of generality, to the antenna
configurations shown in \eqref{eq:gdof-zic-antenna-assumptions-end}.
Henceforth, we operate under the antenna assumptions in \eqref{eq:gdof-zic-antenna-assumptions-end}
for the rest of the paper. 

\emph{\uline{Summary of achievable scheme}}:

The general achievability scheme has a block-Markov structure, which
consists of $B$ blocks, each consisting of $s$ time slots. In this
paper, without loss of generality, we take $s=1$, i.e., each block
consists of a single time slot. In block $b$, each transmitter $T_{i}$,
$i\in\left\{ 1,2\right\} $ transmits a message $w_{i,b}$ intended
for $R_{i}$ using $M_{i}$ data streams%
. While $T_{1}$ transmits its message at full power, $T_{2}$ modulates
its transmit power level, parameterized by $A_{2}$, based on the
GDoF tuple to be achieved. The interference seen at $R_{1}$ in the
previous block $b-1$, which has a power level $\rho^{\left(\alpha-A_{2}\right)}$,
can be reconstructed at $T_{2}$ in this block, using delayed CSIT
of the cross-link $H_{12}$. $T_{2}$ uses digital quantization to
compress this interference, such that the average distortion does
not exceed the noise level (which can then be ignored from a GDoF
perspective) and digitally multicasts the quantization index $l_{b-1}$
with full power, after encoding it as a common message $x_{2c}\left(l_{b-1}\right)$.
This common message is useful at both the receivers while decoding
block $b-1$, since it provides $R_{1}$ with enough information to
subtract an estimate of the interference seen in block $b-1$, and
provides $R_{2}$ useful side-information about its own message $w_{2,b-1}$.
Decoding starts from the final block, proceeding successively backwards
to the first block, using the common message decoded in block $b$
as side-information while decoding block $b-1$. We thus obtain a
general achievability region, parameterized by the power level $A_{2}$.
This allows us to specify the transmit power level $A_{2}$ required
to achieve each non-trivial corner point of the GDoF region from Theorem
\ref{thm:Main-Result}, separately for weak and strong interference
in the next two sections, respectively. 

Because of the similar structure of our achievability scheme with
the DoF achievability scheme for the IC with mixed CSIT in \cite{Yi2014},
we adopt the notation from \cite{Yi2014} for presentation purposes.
But, as explained earlier in the introduction, our subsequent analysis
differs considerably from the DoF analysis in \cite{Yi2014}. 

\emph{\uline{Encoding and transmission strategy}}:

In each block $b$, transmitter $T_{i}$, $i\in\left\{ 1,2\right\} $,
encodes its private message $w_{i,b}$ as the vector $u_{i}\left(w_{i,b}\right)\in\mathbb{C}^{M_{i}\times1}$,
such that $u_{i}\left(w_{i,b}\right)\sim\mathcal{CN}\left(0,Q_{i}\right)$,
where $Q_{1}\triangleq\mathbf{I}_{M_{1}}$ and $Q_{2}\triangleq\rho^{-A_{2}}\mathbf{I}_{M_{2}}$,
with $A_{2}\geq0$. We note that, by using the $\sim$ notation, we
can omit a constant multiplicative factor for the covariance matrices
which does not affect the GDoF analysis. In other words, while $T_{1}$
transmits its message at full power $P_{T_{1}}\sim\rho^{0}$, $T_{2}$
transmits its private message at less than full power $P_{T_{2}}\sim\rho^{-A_{2}}$.
In the next two sections, the power level $A_{2}$ will be specified
separately for each GDoF corner point. $T_{2}$ also encodes the common
message $l_{b-1}$, to be defined later, using the vector $x_{2c}\left(l_{b-1}\right)\in\mathbb{C}^{M_{2}\times1}$
and transmits it with maximum power $P_{c}\sim\rho^{0}$. Thus, the
transmitted signal at transmitters $T_{1}$ and $T_{2}$ in block
$b$ are, respectively as follows:
\begin{eqnarray*}
x_{1}\left[b\right] & = & u_{1}\left(w_{1,b}\right),\\
x_{2}\left[b\right] & = & u_{2}\left(w_{2,b}\right)+x_{2c}\left(l_{b-1}\right).
\end{eqnarray*}
Since there is no common message in the first block $b=1$, we set
$l_{0}=0$. In the final block $b=B$, only the common message is
transmitted, and so we set $w_{1,B}=w_{2,B}=0$, to end the transmission. 

In block $b$, the received signal at each receiver, and the power
level of each constituent of the received signal (indicated below
it), is as follows: %
\begin{eqnarray*}
Y_{1}\left[b\right] & = & \underbrace{\sqrt{\rho^{\alpha}}H_{12}\left[b\right]x_{2c}\left(l_{b-1}\right)}_{\rho^{\alpha}}+\underbrace{\sqrt{\rho}H_{11}\left[b\right]u_{1}\left(w_{1,b}\right)}_{\rho^{1}}+\\
 &  & +\underbrace{\sqrt{\rho^{\alpha}}H_{12}\left[b\right]u_{2}\left(w_{2,b}\right)}_{\eta_{b}\sim\rho^{\left(\alpha-A_{2}\right)}},
\end{eqnarray*}
and

\[
Y_{2}\left[b\right]=\underbrace{\sqrt{\rho}H_{22}\left[b\right]x_{2c}\left(l_{b-1}\right)}_{\rho^{1}}+\underbrace{\sqrt{\rho}H_{22}\left[b\right]u_{2}\left(w_{2,b}\right)}_{\rho^{\left(1-A_{2}\right)}},
\]
where $\eta_{b}$ is the interference caused at $R_{1}$ by $u_{2}$.
This interference $\eta_{b}$ is reconstructed at $T_{2}$ in the
next block $b+1$ through delayed CSIT of the channel $H_{12}$$\left[b\right]$.
From the rate distortion theorem \cite{Cover2006}, we know that this
interference, which is at power level $\rho^{\left(\alpha-A_{2}\right)}$
and carries a GDoF of no more than that of $u_{2}$, i.e., $N_{2}$,
can be quantized at $T_{2}$ using a source codebook of size $\rho^{\min\left(N_{2},\left(\alpha-A_{2}\right)N_{1}^{\prime}\right)}$,
such that the mean square distortion does not exceed the noise level
and can thus be ignored from a GDoF perspective. The quantization
index is denoted as $l_{b}$, and we also define $d_{\eta}$ as the
GDoF carried by the common message. The GDoF carried by the private
message $w_{i,b}$ is denoted as $d_{ib}$, $\forall i\in\left\{ 1,2\right\} $. 

\emph{\uline{Decoding}}:

The decoding starts from the last block $B$, in which both receivers
decode only the common message $l_{B-1}$, since $w_{1,B}=w_{2,B}=0$.
This is possible only when
\begin{eqnarray}
d_{\eta} & \leq & \alpha\min\left(M_{2},N_{1}\right)=\alpha N_{1}^{\prime},\\
d_{\eta} & \leq & \min\left(M_{2},N_{2}\right),
\end{eqnarray}
where the two conditions are the decoding conditions at $R_{1}$ and
$R_{2}$, respectively. The first condition comes from the MIMO point-to-point
channel from $T_{2}$ to $R_{1},$ with the received power of the
common message $l_{B-1}$ being $\rho^{\alpha}$, and similarly, the
second condition comes from the MIMO point-to-point channel from $T_{2}$
to $R_{2}$, where the message $l_{B-1}$ is received at power $\rho$. 

The decoding process now moves backwards to the previous block. Thus,
while decoding block $b$, the common message $l_{b}$ is known at
each receiver from the decoding of block $b+1$. Receiver $R_{1}$
can thus reconstruct the interference $\eta_{b}$ (with a distortion
that does not exceed the noise level which can thus be neglected)
and subtract it from its received signal, while receiver $R_{2}$
reconstructs $\eta_{b}$ to use it as side-information. Now, the signal
at each receiver is that of a $2$-user MAC, as shown below (for the
sake of brevity, we omit the block indices $b$ for the channel matrices):
\begin{align}
Y_{1}\left[b\right]-\eta_{b} & =\sqrt{\rho^{\alpha}}H_{12}x_{2c}\left(l_{b-1}\right)+\sqrt{\rho}H_{11}u_{1}\left(w_{1,b}\right),\label{eq:R1-MAC}\\
\left[\begin{array}{c}
Y_{2}\left[b\right]\\
\eta_{b}
\end{array}\right] & =\negmedspace\negmedspace\left[\begin{array}{c}
\sqrt{\rho}H_{22}\negmedspace\\
0
\end{array}\right]\negmedspace x_{2c}\negmedspace\left(l_{b-1}\right)\negmedspace+\negmedspace\left[\begin{array}{c}
\negmedspace\sqrt{\rho}H_{22}\negmedspace\negmedspace\\
\negmedspace\sqrt{\rho^{\alpha}}H_{12}\negmedspace\negmedspace
\end{array}\right]\negmedspace u_{2}\negmedspace\left(w_{2,b}\right).\nonumber \\
\label{eq:R2-MAC}
\end{align}
In Appendix \ref{sec:Achievable-GDoF-Region}, we prove that each
receiver $R_{i}$, $i\in\left\{ 1,2\right\} $ can decode its own
message $w_{i,b}$, carrying $d_{ib}$ GDoF, as well as the common
message $l_{b-1}$, which carries $d_{\eta}$ GDoF, if the $\left(d_{1b},d_{2b},d_{\eta}\right)$
tuple lies within the following achievable region:
\begin{eqnarray}
d_{\eta} & \leq & \min\left(\alpha N_{1}^{\prime},N_{2}\right),\label{eq:gdof-zic-equation-range-start}\\
d_{1b} & \leq & M_{1},\label{eq:gdof-zic-equation-d1}\\
d_{\eta}+d_{1b} & \leq & f\left(N_{1},\left(\alpha,M_{2}\right),\left(1,M_{1}\right)\right),\label{eq:gdof-zic-equation-d1-deta}\\
d_{2b} & \leq & f\left(M_{2},\left(1-A_{2},N_{2}\right),\left(\alpha-A_{2},N_{1}\right)\right),\label{eq:gdof-zic-equation-d2}\\
d_{\eta}+d_{2b} & \leq & \left(\alpha-A_{2}\right)N_{1}^{\prime}+N_{2}.\label{eq:gdof-zic-equation-d2-deta}
\end{eqnarray}
Inequalities \eqref{eq:gdof-zic-equation-range-start}-\eqref{eq:gdof-zic-equation-d2-deta}
will be referred to as the general achievability conditions. Since
no private messages are transmitted in the final block, the final
achievable GDoF using this achievable scheme is $d_{i}\triangleq\frac{1}{B}\sum_{b=1}^{B-1}d_{ib}=\frac{B-1}{B}d_{ib}$,
$i\in\left\{ 1,2\right\} $, where we have allocated the same $d_{ib}$
in each block $b$, and by taking $B\rightarrow\infty$, we get $d_{i}\rightarrow d_{ib}$.
Henceforth, we replace $d_{1b}$ and $d_{2b}$ with $d_{1}$ and $d_{2}$,
respectively. In the following two sections, we show how this general
achievable scheme can be used to achieve the GDoF outer bound region
in both the weak and strong interference regimes.

\section{WEAK INTERFERENCE - ACHIEVABILITY\label{sec:WEAK-INTERFERENCE}}

In the weak interference regime $\left(\alpha\leq1\right)$, the general
achievability conditions from the previous section can be simplified
using the following lemma. 
\begin{lem}
\label{lem:Weak-int}When $\alpha\leq1$, the following $\left(d_{1},d_{2}\right)$
GDoF tuple can be achieved: 
\begin{align}
d_{1} & \triangleq\min\left(M_{1},M_{1}-\alpha\left(M_{1}+N_{1}^{\prime}-N_{1}\right)+N_{1}^{\prime}A_{2}\right),\label{eq:weak-int-lemma-d1}\\
d_{2} & \triangleq\min\left(N_{2},N_{2}+\alpha\left(M_{2}-N_{2}\right)-M_{2}A_{2}\right),\label{eq:weak-int-lemma-d2}
\end{align}
whenever $A_{2}$, defined in Section \ref{sec:ACHIEVABLE-SCHEME},
lies in the range 
\begin{equation}
\alpha\geq A_{2}\geq\left(\alpha-\frac{N_{2}}{N_{1}^{\prime}}\right)^{+}.\label{eq:A2-weak-int-range}
\end{equation}
\end{lem}
\begin{IEEEproof}
To prove the lemma, we need to show that the $\left(d_{1},d_{2}\right)$
tuple defined above satisfies the general achievability conditions
\eqref{eq:gdof-zic-equation-range-start}-\eqref{eq:gdof-zic-equation-d2-deta}.
We allocate 
\begin{equation}
d_{\eta}\triangleq\left(\alpha-A_{2}\right)N_{1}^{\prime},\label{eq:weak-int-deta-assigned}
\end{equation}
which makes the first general achievability condition \eqref{eq:gdof-zic-equation-range-start}
redundant, as long as $A_{2}$ lies in the range shown in \eqref{eq:A2-weak-int-range}.
After substituting \eqref{eq:weak-int-deta-assigned} in the remaining
achievability conditions \eqref{eq:gdof-zic-equation-d1}-\eqref{eq:gdof-zic-equation-d2-deta},
we simplify the terms involving $f\negmedspace\left(\right)$, by
using $\alpha\leq1$. The general achievability conditions \eqref{eq:gdof-zic-equation-d1}
and \eqref{eq:gdof-zic-equation-d1-deta} are thus respectively simplified
as follows: 
\begin{eqnarray}
d_{1} & \leq & M_{1},\label{eq:weak-int-proof-d1-1}\\
d_{1} & \leq & M_{1}+\alpha\left(N_{1}-M_{1}\right)-\left(\alpha-A_{2}\right)N_{1}^{\prime}.\label{eq:weak-int-proof-d1-2}
\end{eqnarray}
Similarly, the simplified general achievability conditions \eqref{eq:gdof-zic-equation-d2}
and \eqref{eq:gdof-zic-equation-d2-deta} are, respectively, 
\begin{eqnarray}
d_{2} & \leq & \left(1-A_{2}\right)N_{2}+\left(\alpha-A_{2}\right)\left(M_{2}-N_{2}\right),\label{eq:weak-int-proof-d2-1}\\
d_{2} & \leq & N_{2}.\label{eq:weak-int-proof-d2-3}
\end{eqnarray}

Now, by combining \eqref{eq:weak-int-proof-d1-1} and \eqref{eq:weak-int-proof-d1-2},
it is clear that $d_{1}$ defined in \eqref{eq:weak-int-lemma-d1}
satisfies the general achievability conditions, while achievability
of $d_{2}$ defined in \eqref{eq:weak-int-lemma-d2} similarly follows
from \eqref{eq:weak-int-proof-d2-1} and \eqref{eq:weak-int-proof-d2-3},
and the lemma is thus proved.
\end{IEEEproof}
Using the above lemma, we show that the outer bound region from Theorem
\ref{thm:Main-Result} is achievable when $\alpha<1$. In the weak
interference regime, under the antenna assumptions \eqref{eq:gdof-zic-antenna-assumptions-end},
the outer bound region from Theorem \ref{thm:Main-Result} is given
by the following set of inequalities: 
\begin{eqnarray}
d_{1} & \leq & M_{1},\label{eq:weak-int-bound-PTP}\\
d_{2} & \leq & N_{2},\\
\frac{d_{1}}{N_{1}^{\prime}}+\frac{d_{2}}{M_{2}} & \leq & \negthickspace\negmedspace\negmedspace\frac{M_{1}+\left(N_{1}-M_{1}\right)\alpha}{N_{1}^{\prime}}\negmedspace+\negmedspace\frac{N_{2}+\left(M_{2}-N_{2}\right)\alpha}{M_{2}}\negmedspace-\alpha.\nonumber \\
\label{eq:weak-int-bound-del-csit}
\end{eqnarray}
This leads to two possible shapes of the outer bound region, depending
on whether the delayed CSIT bound \eqref{eq:weak-int-bound-del-csit}
is active or not. We analyze the two cases below, and show that the
GDoF region is achievable in both cases. 
\begin{figure}[tb]
\includegraphics[scale=0.5]{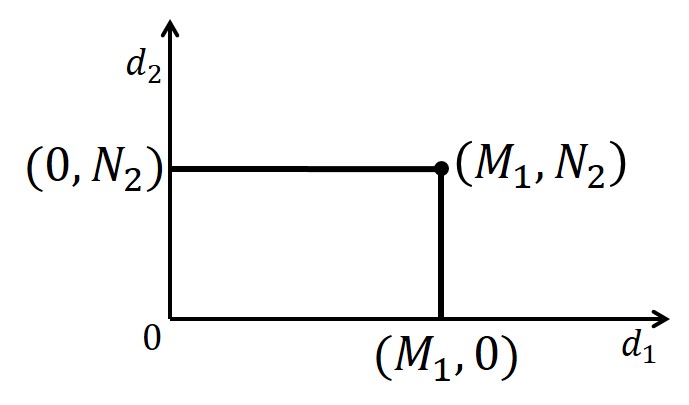}\caption{\label{fig:GDoF-region-Case-I}GDoF region of the MIMO Z-IC with delayed
CSIT, when bound \eqref{eq:delayed-CSIT-bound} is inactive, as seen
in Case I of Sections \ref{sec:WEAK-INTERFERENCE} and \ref{sec:STRONG-INTERFERENCE}.}

\end{figure}
\begin{figure}[tb]
\includegraphics[scale=0.5]{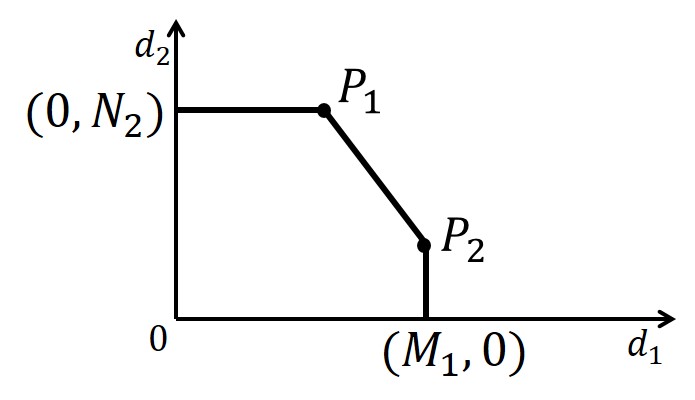}\caption{\label{fig:GDoF-region-Case-II}GDoF region of the MIMO Z-IC with
delayed CSIT, when bound \eqref{eq:delayed-CSIT-bound} is active.
The corner points $P_{1}$ and $P_{2}$ are defined separately for
weak and strong interference in Case II of Sections \ref{sec:WEAK-INTERFERENCE}
and \ref{sec:STRONG-INTERFERENCE}, respectively.}
\end{figure}

\emph{Case I) when $\frac{N_{1}-M_{1}}{N_{1}^{\prime}}\geq\frac{N_{2}}{M_{2}}$:}

In this case, the delayed CSIT bound \eqref{eq:weak-int-bound-del-csit}
is inactive, and the corresponding GDoF outer bound region is shown
in Fig. \ref{fig:GDoF-region-Case-I}. The only non-trivial GDoF corner
point is $\left(M_{1,}N_{2}\right)$, which can be achieved by setting
the transmission power level at $T_{2}$ to 
\begin{equation}
A_{2}=\left(1-\frac{N_{2}}{M_{2}}\right)\alpha,\label{eq:A2-case1-weak-int}
\end{equation}
which satisfies condition \eqref{eq:A2-weak-int-range} in Lemma \ref{lem:Weak-int}.
Substituting \eqref{eq:A2-case1-weak-int} in \eqref{eq:weak-int-lemma-d1}
from Lemma \ref{lem:Weak-int}, we see that the GDoF achieved by the
first user is
\[
d_{1}=\min\left(M_{1},M_{1}+\alpha N_{1}^{\prime}\left(\frac{N_{1}-M_{1}}{N_{1}^{\prime}}-\frac{N_{2}}{M_{2}}\right)\right)=M_{1}.
\]
Similarly, substituting \eqref{eq:A2-case1-weak-int} in \eqref{eq:weak-int-lemma-d2},
we see that $d_{2}=N_{2}$ is achievable, thus proving that the GDoF
tuple $\left(M_{1},N_{2}\right)$ is achievable.

\emph{Case II) when $\frac{N_{1}-M_{1}}{N_{1}^{\prime}}<\frac{N_{2}}{M_{2}}$:}

In this case, the delayed CSIT bound \eqref{eq:weak-int-bound-del-csit}
is active, and the GDoF outer bound region is shown in Fig. \ref{fig:GDoF-region-Case-II},
which contains two non-trivial corner points, which are listed below:
\begin{equation}
\begin{array}{c}
P_{1}\triangleq\left(M_{1}-\alpha\left(M_{1}+N_{1}^{\prime}-N_{1}\right)+\frac{N_{1}^{\prime}}{M_{2}}\left(M_{2}-N_{2}\right)\alpha,\;N_{2}\right),\\
P_{2}\triangleq\left(M_{1},\;N_{2}+\alpha\left(M_{2}-N_{2}\right)-\frac{M_{2}}{N_{1}^{\prime}}\left(N_{1}^{\prime}-N_{1}+M_{1}\right)\alpha\right).
\end{array}
\end{equation}
The first corner point $P_{1}$ is achieved by setting the transmission
power level as follows: 
\begin{equation}
A_{2}=\left(1-\frac{N_{2}}{M_{2}}\right)\alpha,\label{eq:A2-case2-weak-int-P1}
\end{equation}
and the second corner point $P_{2}$ is achieved with the transmission
power level 
\begin{equation}
A_{2}=\left(1-\frac{N_{1}-M_{1}}{N_{1}^{\prime}}\right)\alpha.\label{eq:A2-case2-weak-int-P2}
\end{equation}
For both the points, the assigned value of $A_{2}$ satisfies the
condition \eqref{eq:A2-weak-int-range} in Lemma \ref{lem:Weak-int},
and it is straightforward to check that by substituting \eqref{eq:A2-case2-weak-int-P1}
and \eqref{eq:A2-case2-weak-int-P2} in Lemma \ref{lem:Weak-int},
we achieve the points $P_{1}$ and $P_{2}$, respectively.

\subsection{Comparison with DoF region}

The DoF region of the MIMO Z-IC with delayed CSIT (with the antenna
conditions \eqref{eq:gdof-zic-antenna-assumptions-end}) was obtained
in \cite{Mohanty2015} and is given below: 
\begin{eqnarray}
d_{1} & \leq & M_{1},\nonumber \\
d_{2} & \leq & N_{2},\nonumber \\
d_{1}+d_{2} & \leq & \max\left(M_{2},N_{1}\right),\nonumber \\
\frac{d_{1}}{N_{1}^{\prime}}+\frac{d_{2}}{M_{2}} & \leq & \frac{N_{1}}{N_{1}^{\prime}}.\label{eq:weak-int-bound-1}\\
\nonumber 
\end{eqnarray}
Comparing it with the GDoF region with weak interference shown in
\eqref{eq:weak-int-bound-PTP}-\eqref{eq:weak-int-bound-del-csit},
it is straightforward to check that the DoF delayed CSIT region \eqref{eq:weak-int-bound-1}
is always smaller than or identical to the corresponding GDoF delayed
CSIT region for all values of $\alpha\leq1$.

\subsection{Comparison with perfect CSIT}

The GDoF region of the MIMO IC with perfect CSIT was obtained in \cite{Karmakar2012},
from which, by setting $\alpha_{11}=\alpha_{22}=1$, $\alpha_{12}=0$
and $\alpha_{21}=\alpha$ in Theorem $1$ therein, we obtain the GDoF
region of the MIMO Z-IC with perfect CSIT, given below (under the
antenna conditions \eqref{eq:gdof-zic-antenna-assumptions-end}):
\begin{align}
d_{1}\leq M_{1,} & \qquad d_{2}\leq N_{2},\nonumber \\
d_{1}+d_{2} & \leq f\left(N_{1},\left(\alpha,M_{2}\right),\left(1,M_{1}\right)\right)\nonumber \\
 & +f\left(N_{2},\left(1-\alpha,N_{1}^{\prime}\right),\left(1,M_{2}-N_{1}^{\prime}\right)\right).\label{eq:perfect-CSIT-GDoF-region}
\end{align}

When $\alpha\leq1$ and $M_{2}\leq N_{1}$, i.e., $N_{1}^{\prime}=M_{2}$,
it is straightforward to show that the GDoF region with only delayed
CSIT, shown in \eqref{eq:weak-int-bound-PTP}-\eqref{eq:weak-int-bound-del-csit},
coincides with the perfect CSIT GDoF region in \eqref{eq:perfect-CSIT-GDoF-region}.

When $M_{2}>N_{1}$ and $\alpha\leq1$, the non-trivial perfect CSIT
bound in \eqref{eq:perfect-CSIT-GDoF-region} is always loose compared
to the corresponding delayed CSIT bound \eqref{eq:weak-int-bound-del-csit}.
Thus, when $M_{2}>N_{1}$ and the delayed CSIT bound \eqref{eq:weak-int-bound-del-csit}
is active, i.e., when \emph{$\frac{N_{1}-M_{1}}{N_{1}}<\frac{N_{2}}{M_{2}}$}
(Case II above), the delayed CSIT GDoF region with weak interference
is strictly smaller than the corresponding perfect CSIT GDoF region.
Otherwise, for antenna configurations from Case I with $M_{2}>N_{1}$,
the delayed CSIT and perfect CSIT GDoF regions coincide for $\alpha\leq1$.

\subsection{Sum-GDoF}

For Case I above, we see from the shape of the GDoF region in Fig.
\ref{fig:GDoF-region-Case-I} that the sum-GDoF is 
\[
d_{\Sigma}=M_{1}+N_{2}.
\]
For Case II, the straight line in the $\left(d_{1},d_{2}\right)$
plane that defines the outer bound \eqref{eq:delayed-CSIT-bound}
has slope $-\frac{M_{2}}{N_{1}^{\prime}}<-1$, and thus, the maximum
sum-GDoF is achieved at the corner point $P_{1}$ in Fig. \ref{fig:GDoF-region-Case-II},
and is equal to 
\[
d_{\Sigma}=M_{1}+N_{2}-\alpha\left(M_{1}-N_{1}+\frac{N_{2}N_{1}^{\prime}}{M_{2}}\right),
\]
which decreases linearly with $\alpha$ in the weak interference regime.
This proves the first part \eqref{eq:sum-GDoF-weak-int} of Corollary
\ref{cor:sum-GDoF}.

For comparison, the sum-GDoF with perfect CSIT can be obtained from
\eqref{eq:perfect-CSIT-GDoF-region}. When $\alpha\leq1$ and
$M_{2}>N_{1}$ (since the perfect CSIT and delayed CSIT GDoF regions
are the same when $M_{2}\leq N_{1}$), the perfect CSIT sum-GDoF is
as follows: 
\[
d_{\sum}^{\mbox{p}}=\begin{cases}
M_{1}+N_{2}, & M_{1}\negthinspace+\negthinspace N_{2}\leq M_{2}\\
M_{1}\left(1-\alpha\right)+M_{2}\alpha+N_{2}\left(1-\alpha\right), & M_{1}\negthinspace+\negthinspace N_{2}>M_{2}.
\end{cases}
\]

\section{STRONG INTERFERENCE - ACHIEVABILITY\label{sec:STRONG-INTERFERENCE}}

When $\alpha>1$, the general achievability conditions \eqref{eq:gdof-zic-equation-range-start}-\eqref{eq:gdof-zic-equation-d2-deta}
can be simplified using the following lemma. 
\begin{lem}
\label{lem:strong-int}When $\alpha>1$, the general achievability
scheme can achieve the GDoF tuple $\left(d_{1},d_{2}\right)$ shown
below: 
\begin{equation}
d_{1}\triangleq\begin{cases}
\min\left(M_{1},\left(\alpha-1\right)N_{1}^{\prime}+\left(N_{1}-N_{2}\right)\right), & A_{2}<\alpha-\frac{N_{2}}{N_{1}^{\prime}}\\
\min\left(M_{1},N_{1}-N_{1}^{\prime}+N_{1}^{\prime}A_{2}\right), & A_{2}\geq\alpha-\frac{N_{2}}{N_{1}^{\prime}}
\end{cases}\label{eq:d1-assign-strong-int}
\end{equation}
and
\begin{equation}
d_{2}\triangleq\begin{cases}
N_{2},\mathrm{\qquad\qquad\qquad\qquad\qquad\qquad\quad}A_{2}<\alpha-\frac{N_{2}}{N_{1}^{\prime}}\\
\min\negmedspace\left(\negmedspace N_{2},\left(\alpha\negmedspace-\negmedspace A_{2}\right)N_{1}^{\prime}+\left(1\negmedspace-\negmedspace A_{2}\right)^{+}\left(M_{2}\negmedspace-\negmedspace N_{1}^{\prime}\right)\right),\\
\mathrm{\qquad\qquad\qquad\qquad\qquad\qquad\qquad\quad}A_{2}\geq\alpha-\frac{N_{2}}{N_{1}^{\prime}},
\end{cases}\label{eq:d2-assign-strong-int}
\end{equation}
where $A_{2}$ is defined in Section \ref{sec:ACHIEVABLE-SCHEME}.\end{lem}
\begin{IEEEproof}
In the strong interference regime, we assign the value of $d_{\eta}$
as follows:
\begin{equation}
d_{\eta}\triangleq\begin{cases}
N_{2}, & A_{2}<\alpha-\frac{N_{2}}{N_{1}^{\prime}}\\
\left(\alpha-A_{2}\right)N_{1}^{\prime}, & A_{2}\geq\alpha-\frac{N_{2}}{N_{1}^{\prime}},
\end{cases}\label{eq:strong-int-deta-assign}
\end{equation}
which trivially satisfies the first general achievability condition
\eqref{eq:gdof-zic-equation-range-start}. Moreover, since $\alpha>1$,
the general achievability condition \eqref{eq:gdof-zic-equation-d1-deta}
simplifies to 
\[
d_{\eta}+d_{1}\leq\alpha N_{1}^{\prime}+\left(N_{1}-N_{1}^{\prime}\right).
\]
Substituting the value of $d_{\eta}$ from \eqref{eq:strong-int-deta-assign}
in the above condition, and combining it with the achievability condition
\eqref{eq:gdof-zic-equation-d1}, i.e., $d_{1}\leq M_{1}$, we see
that $d_{1}$ shown in \eqref{eq:d1-assign-strong-int} is always
achievable.

Since $\alpha>1$, the general achievability condition \eqref{eq:gdof-zic-equation-d2}
simplifies to 
\begin{equation}
d_{2}\leq\left(\alpha-A_{2}\right)N_{1}^{\prime}+\left(1-A_{2}\right)^{+}\left(M_{2}-N_{1}^{\prime}\right).\label{eq:proof-strong-int-d2-cond1}
\end{equation}
Moreover, when $A_{2}\geq\alpha-\frac{N_{2}}{N_{1}^{\prime}}$, we
also have the condition 
\begin{equation}
d_{2}\leq N_{2},\label{eq:proof-strong-int-d2-cond2}
\end{equation}
 obtained by substituting $d_{\eta}=\left(\alpha-A_{2}\right)N_{1}^{\prime}$
in the remaining general achievability condition \eqref{eq:gdof-zic-equation-d2-deta}.
Combining \eqref{eq:proof-strong-int-d2-cond1} and \eqref{eq:proof-strong-int-d2-cond2},
it is clear that when $A_{2}\geq\alpha-\frac{N_{2}}{N_{1}^{\prime}}$,
the value of $d_{2}$ shown in \eqref{eq:d2-assign-strong-int} is
achievable. 

When $A_{2}<\alpha-\frac{N_{2}}{N_{1}^{\prime}}$, i.e., when $d_{\eta}=N_{2}$,
the general achievability conditions \eqref{eq:gdof-zic-equation-d2}
and \eqref{eq:gdof-zic-equation-d2-deta} can be reframed, respectively,
as \eqref{eq:proof-strong-int-d2-cond1} and 
\[
d_{2}\leq\left(\alpha-A_{2}\right)N_{1}^{\prime},
\]
both of which are satisfied by assigning $d_{2}=N_{2}$, since we
have $N_{2}\leq\left(\alpha-A_{2}\right)N_{1}^{\prime}$ in this case.
Thus, the value of $d_{2}$ specified in the lemma is always achievable,
proving the lemma.\end{IEEEproof}
\begin{rem}
\label{rem:A2d2}The achievable $d_{2}$ in \eqref{eq:d2-assign-strong-int}
from Lemma \ref{lem:strong-int} is clearly a monotonically decreasing
function of $A_{2}$, and with some straightforward algebraic manipulation,
it is easily shown that the value of $d_{2}$ in \eqref{eq:d2-assign-strong-int}
remains constant at $N_{2}$ when $A_{2}\leq A_{2}^{d2}$, and strictly
monotonically decreases with $A_{2}$ when $A_{2}>A_{2}^{d2}$, where
$A_{2}^{d2}$ is given below: 
\begin{equation}
A_{2}^{d2}\triangleq\begin{cases}
1+\frac{\left(\alpha-1\right)N_{1}^{\prime}-N_{2}}{M_{2}}, & \alpha<1+\frac{N_{2}}{N_{1}^{\prime}}\\
\alpha-\frac{N_{2}}{N_{1}^{\prime}}, & \alpha\geq1+\frac{N_{2}}{N_{1}^{\prime}}.
\end{cases}\label{eq:gdof-zic-strong-int-A2d2-lim}
\end{equation}

\end{rem}
In the strong interference regime, and using the antennas assumptions
in \eqref{eq:gdof-zic-antenna-assumptions-end}, the outer bound region
from Theorem \ref{thm:Main-Result} is as follows: 
\begin{eqnarray}
d_{1} & \leq & M_{1},\nonumber \\
d_{2} & \leq & N_{2},\nonumber \\
\frac{d_{1}}{N_{1}^{\prime}}+\frac{d_{2}}{M_{2}} & \leq & \frac{N_{1}}{N_{1}^{\prime}}+\frac{\left(\alpha-1\right)N_{1}^{\prime}}{M_{2}},\label{eq:strong-int-bound}
\end{eqnarray}
where we have used the following simplifications for strong interference,
\[
\begin{array}{c}
f\left(N_{1},\left(\alpha,M_{2}\right),\left(1,M_{1}\right)\right)=\alpha N_{1}^{\prime}+\left(N_{1}-N_{1}^{\prime}\right),\\
f\left(M_{2},\left(\alpha,N_{1}\right),\left(1,N_{2}\right)\right)=\alpha N_{1}^{\prime}+M_{2}-N_{1}^{\prime}.
\end{array}
\]
Similar to the weak interference regime, the above GDoF outer bound
region can have two different shapes, as shown below, depending on
whether the delayed CSIT bound \eqref{eq:strong-int-bound} is active
or not. We now show that the GDoF outer bound region with strong interference
$\left(\alpha>1\right)$ is achievable in both cases, shown below: 

\emph{Case I) when $\alpha\geq1+\frac{N_{2}}{N_{1}^{\prime}}-\frac{M_{2}\left(N_{1}-M_{1}\right)}{\left(N_{1}^{\prime}\right)^{2}}$:}

In this case, the delayed CSIT bound \eqref{eq:strong-int-bound}
is inactive, and the GDoF outer bound region is shown in Fig. \ref{fig:GDoF-region-Case-I}.
The only non-trivial GDoF corner point is $\left(M_{1,}N_{2}\right)$,
which can be achieved by setting the transmission power level at $T_{2}$
as follows: 
\begin{equation}
A_{2}=1-\frac{N_{1}-M_{1}}{N_{1}^{\prime}}.\label{eq:strong-int-case-I-A2}
\end{equation}

To see that the assigned value of $A_{2}$ achieves $d_{2}=N_{2}$,
it suffices to show that $A_{2}\leq A_{2}^{d2}$, as per Remark \ref{rem:A2d2}.
To this end, we consider both sub-cases of \eqref{eq:gdof-zic-strong-int-A2d2-lim}.
When $\alpha<1+\frac{N_{2}}{N_{1}^{\prime}}$, we see that $A_{2}\leq A_{2}^{d2}$,
as shown below: 
\begin{eqnarray*}
\alpha & \geq & 1+\frac{N_{2}}{N_{1}^{\prime}}-\frac{M_{2}\left(N_{1}-M_{1}\right)}{\left(N_{1}^{\prime}\right)^{2}}\\
\Rightarrow1-\frac{N_{1}-M_{1}}{N_{1}^{\prime}} & \leq & 1+\frac{\left(\alpha-1\right)N_{1}^{\prime}-N_{2}}{M_{2}}.
\end{eqnarray*}
When $\alpha\geq1+\frac{N_{2}}{N_{1}^{\prime}}$, we have 
\begin{eqnarray*}
\alpha & \geq & 1+\frac{N_{2}}{N_{1}^{\prime}}-\frac{N_{1}-M_{1}}{N_{1}^{\prime}}\\
\Rightarrow1-\frac{N_{1}-M_{1}}{N_{1}^{\prime}} & \leq & \alpha-\frac{N_{2}}{N_{1}^{\prime}},
\end{eqnarray*}
i.e., $A_{2}\leq A_{2}^{d2}$.

Substituting the value of $A_{2}$ in \eqref{eq:d1-assign-strong-int}
achieves $d_{1}=M_{1}$, as we show for each of the two sub-cases
in \eqref{eq:d1-assign-strong-int}. For the first sub-case, by substituting
\eqref{eq:strong-int-case-I-A2} in the condition $A_{2}\negmedspace<\negmedspace\alpha-\frac{N_{2}}{N_{1}^{\prime}}$,
we obtain
\[
M_{1}<\left(\alpha-1\right)N_{1}^{\prime}+N_{1}-N_{2},
\]
and thus, using the above inequality in \eqref{eq:d1-assign-strong-int},
it is clear that $d_{1}=M_{1}$ is achievable. For the second sub-case,
i.e., when $A_{2}\geq\alpha-\frac{N_{2}}{N_{1}^{\prime}}$, we see
that 
\[
N_{1}-N_{1}^{\prime}+N_{1}^{\prime}A_{2}=N_{1}-N_{1}^{\prime}+N_{1}^{\prime}\left(1-\frac{N_{1}-M_{1}}{N_{1}^{\prime}}\right)=M_{1}
\]
 and thus $d_{1}=M_{1}$ is achievable for this sub-case too.

\emph{Case II: when $\alpha<1+\frac{N_{2}}{N_{1}^{\prime}}-\frac{M_{2}\left(N_{1}-M_{1}\right)}{\left(N_{1}^{\prime}\right)^{2}}$:}

In this case, the delayed CSIT bound \eqref{eq:strong-int-bound}
is active, and the GDoF region is shown in Fig. \ref{fig:GDoF-region-Case-II},
where the two non-trivial corner points are as follows:
\[
\begin{array}{c}
P_{1}\triangleq\left(N_{1}+\frac{N_{1}^{\prime}}{M_{2}}\left(\left(\alpha-1\right)N_{1}^{\prime}-N_{2}\right),\;N_{2}\right),\\
P_{2}\triangleq\left(M_{1},\;\left(\alpha-1\right)N_{1}^{\prime}+\frac{M_{2}}{N_{1}^{\prime}}\left(N_{1}-M_{1}\right)\right).
\end{array}
\]

\emph{\uline{Point $P_{1}$}}

The corner point $P_{1}$ is achieved by setting the transmission
power level as 
\begin{equation}
A_{2}=1-\frac{N_{2}-\left(\alpha-1\right)N_{1}^{\prime}}{M_{2}}.\label{eq:strong-int-P1-power-allocation}
\end{equation}
Since, $\alpha<1+\frac{N_{2}}{N_{1}^{\prime}}$ (from the defining
condition of Case II), we see that $A_{2}=A_{2}^{d2}$ (from \eqref{eq:gdof-zic-strong-int-A2d2-lim}
and \eqref{eq:strong-int-P1-power-allocation}) and thus $d_{2}=N_{2}$
is achievable, as per Remark \ref{rem:A2d2}. As for $d_{1}$, we
see that only the second sub-case of \eqref{eq:d1-assign-strong-int}
in Lemma \ref{lem:strong-int} is involved, since $A_{2}\leq\alpha-\frac{N_{2}}{N_{1}^{\prime}}$,
as seen below: 
\begin{align*}
\alpha & <1+\frac{N_{2}}{N_{1}^{\prime}}\\
\Rightarrow\alpha\left(1-\frac{N_{1}^{\prime}}{M_{2}}\right) & \leq\left(1+\frac{N_{2}}{N_{1}^{\prime}}\right)\left(1-\frac{N_{1}^{\prime}}{M_{2}}\right)\\
\Rightarrow1-\frac{N_{2}-\left(\alpha-1\right)N_{1}^{\prime}}{M_{2}} & \leq\alpha-\frac{N_{2}}{N_{1}^{\prime}}.
\end{align*}
Now, substituting the value of $A_{2}$ in the second sub-case of
\eqref{eq:d1-assign-strong-int}, we see that $P_{1}$ is achievable.

\emph{\uline{Point $P_{2}$}}

To achieve the second corner point $P_{2}$, we set the transmission
power level as 
\begin{equation}
A_{2}=1-\frac{N_{1}-M_{1}}{N_{1}^{\prime}}.\label{eq:strong-int-P2-power-allocation}
\end{equation}
It is easy to show that $A_{2}>\alpha-\frac{N_{2}}{N_{1}^{\prime}}$,
as follows: 
\begin{align*}
\alpha & <1+\frac{N_{2}}{N_{1}^{\prime}}-\frac{M_{2}\left(N_{1}-M_{1}\right)}{\left(N_{1}^{\prime}\right)^{2}}\\
\overset{\left(a\right)}{\Rightarrow}\alpha & <1+\frac{N_{2}}{N_{1}^{\prime}}-\frac{N_{1}-M_{1}}{N_{1}^{\prime}}\\
\Rightarrow\alpha-\frac{N_{2}}{N_{1}^{\prime}} & <1-\frac{N_{1}-M_{1}}{N_{1}^{\prime}},
\end{align*}
where $\left(a\right)$ holds true because $M_{2}\geq N_{1}^{\prime}$.
Thus, for both \eqref{eq:d1-assign-strong-int} and \eqref{eq:d2-assign-strong-int}
in Lemma \ref{lem:strong-int}, only the second sub-case is active.
By substituting the value of $A_{2}$ from \eqref{eq:strong-int-P2-power-allocation}
in \eqref{eq:d1-assign-strong-int} and \eqref{eq:d2-assign-strong-int},
it is straightforward to show that $P_{2}$ is achievable.

\subsection{Comparison with DoF region}

Comparing the delayed CSIT DoF region \eqref{eq:weak-int-bound-del-csit}
with the corresponding delayed CSIT GDoF region \eqref{eq:strong-int-bound},
it is clear that the GDoF region with strong interference $\left(\alpha>1\right)$
is always equal to or larger than the corresponding DoF region.

\subsection{Comparison with perfect CSIT}

When $M_{2}\leq N_{1}$ i.e., $N_{1}^{\prime}=M_{2}$, it is straightforward
to show that the delayed CSIT GDoF region shown in \eqref{eq:weak-int-bound-del-csit},
coincides with the perfect CSIT GDoF region in \eqref{eq:perfect-CSIT-GDoF-region}
when $\alpha>1$. 

When $M_{2}>N_{1}$ and $\alpha>1$, the non-trivial perfect CSIT
bound in \eqref{eq:perfect-CSIT-GDoF-region} is always loose compared
to the delayed CSIT bound \eqref{eq:strong-int-bound}. Thus,
when $M_{2}>N_{1}$ and the delayed CSIT bound \eqref{eq:weak-int-bound-del-csit}
is active, i.e., when \emph{$1\negmedspace<\negmedspace\alpha\negmedspace<\negmedspace1\negmedspace+\negmedspace\frac{N_{2}}{N_{1}}\negmedspace-\negmedspace\frac{M_{2}\left(N_{1}-M_{1}\right)}{\left(N_{1}\right)^{2}}$}
(Case II above), the delayed CSIT GDoF region is strictly smaller
than the corresponding perfect CSIT GDoF region. Otherwise, for antenna
configurations from Case I with $M_{2}>N_{1}$, the delayed CSIT and
perfect CSIT GDoF regions coincide for $\alpha>1$.

\subsection{Sum-GDoF}

For Case I above, we see from the shape of the GDoF region in Fig.
\ref{fig:GDoF-region-Case-I} that the maximum sum-GDoF is given by
\[
d_{\Sigma}=M_{1}+N_{2}.
\]
For Case II, the bound \eqref{eq:strong-int-bound} is active. The
straight line in the $\left(d_{1},d_{2}\right)$ plane that defines
this bound has slope $-\frac{M_{2}}{N_{1}^{\prime}}\leq-1$, and thus,
the maximum sum-GDoF in this case is achieved at the corner point
$P_{1}$ in Fig. \ref{fig:GDoF-region-Case-II}, and is equal to 
\[
d_{\Sigma}=\left[N_{2}+N_{1}-\frac{\left(N_{2}+N_{1}^{\prime}\right)N_{1}^{\prime}}{M_{2}}\right]+\frac{\left(N_{1}^{\prime}\right)^{2}}{M_{2}}\alpha
\]
which is an increasing linear function of $\alpha$. This proves the
second and remaining part \eqref{eq:sum-GDoF-strong-int} of Corollary
\ref{cor:sum-GDoF}.

For comparison, we also provide the sum-GDoF with perfect CSIT in
the strong interference regime when $M_{2}>N_{1}$ (recall that the
GDoF regions for delayed CSIT and perfect CSIT coincide for strong
interference when $M_{2}\leq N_{1}$), which, from \eqref{eq:perfect-CSIT-GDoF-region},
we find to be as follows: 
\[
d_{\sum}^{\mbox{p}}=\begin{cases}
M_{1}+N_{2}, & \alpha\geq1+\frac{N_{2}}{N_{1}}-\frac{M_{2}-M_{1}}{N_{1}}\\
M_{2}+\left(\alpha-1\right)N_{1}, & \alpha<1+\frac{N_{2}}{N_{1}}-\frac{M_{2}-M_{1}}{N_{1}}.
\end{cases}
\]

\section{DISCUSSION OF RESULTS \label{sec:INSIGHT}}

\subsection{GDoF vs DoF}

\begin{figure}[tb]
\includegraphics[scale=0.5]{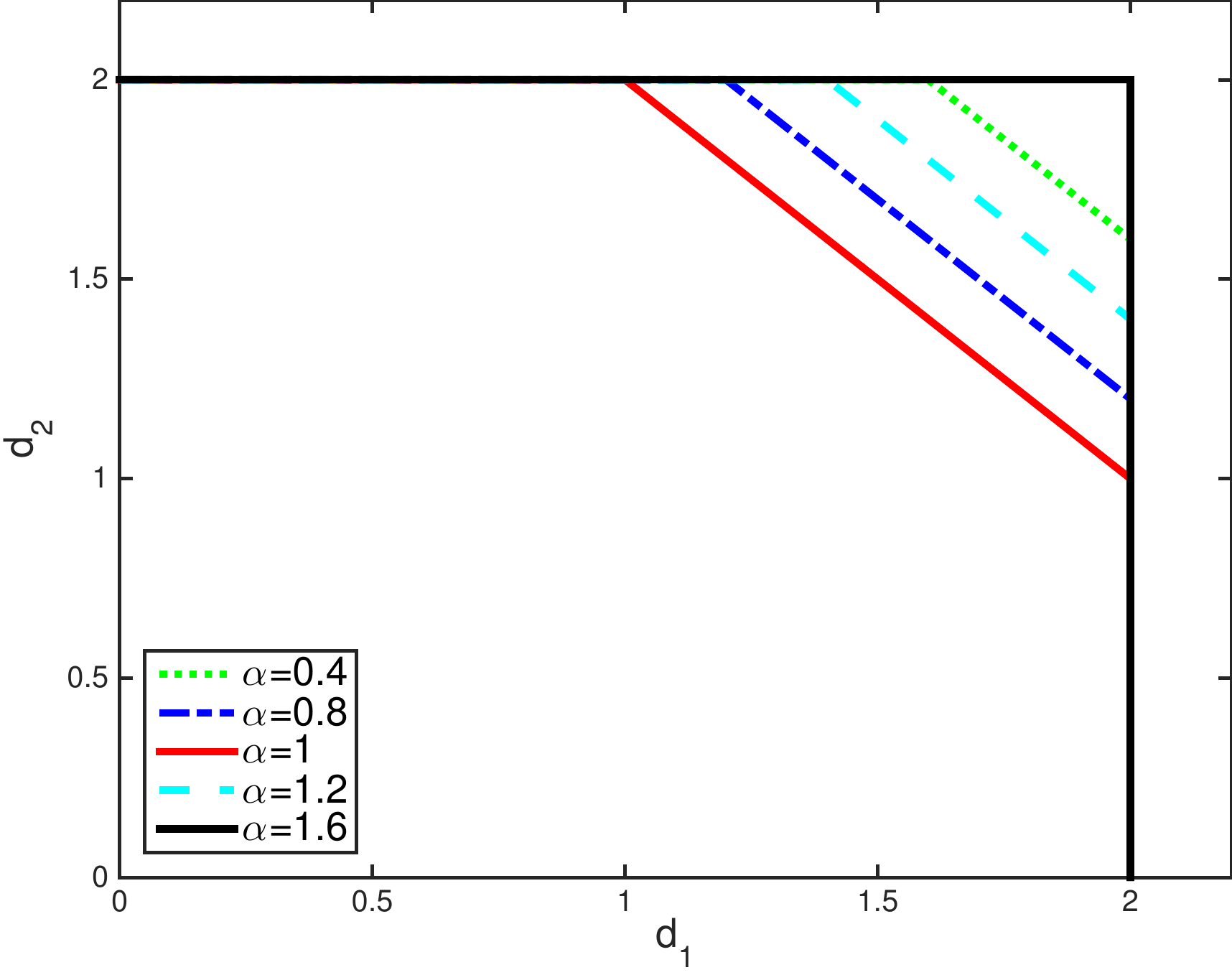}

\caption{\label{fig:GDoF-vs-DoF-comparison}The GDoF region of the $\left(2,2,3,2\right)$
Z-IC at various $\alpha$.}
\end{figure}
When the cross-link of the Z-IC differs in strength from the direct
links, this knowledge about the channel statistics is incorporated
in the achievability scheme developed in this paper. In such a situation,
naively applying the existing DoF-optimal achievability scheme from
\cite{Mohanty2015}, which incorrectly assumes all three links to
be of equal strength, can lead to an achievable region that is strictly
sub-optimal. To demonstrate the benefit of incorporating the channel
statistics into the achievable scheme through a GDoF analysis, we
compared the DoF region $\left(\alpha=1\right)$ of the MIMO Z-IC
with the corresponding GDoF regions in the weak and strong interference
regimes in Sections \ref{sec:WEAK-INTERFERENCE} and \ref{sec:STRONG-INTERFERENCE},
respectively, and showed that the GDoF region, irrespective of the
interference regime, is always larger than or equal to the corresponding
DoF region, for all antenna configurations. In general, when the delayed
CSIT bound \eqref{eq:delayed-CSIT-bound} is active, the delayed CSIT
GDoF region becomes smaller as $\alpha$ increases from 0 to 1, with
the DoF region (at $\alpha=1$) being the smallest, and then,
as $\alpha$ increases for $\alpha>1$, the GDoF region becomes larger,
until the delayed CSIT bound becomes inactive. This is illustrated
for the $\left(2,2,3,2\right)$ Z-IC in Fig. \ref{fig:GDoF-vs-DoF-comparison},
which also shows that the GDoF regions for $\alpha=0.4,\:0.8$ (weak
interference) and $\alpha=1.2,\:1.6$ (strong interference) are strictly
larger than the DoF region $\left(\alpha=1\right)$. 
\begin{figure}[tb]
\includegraphics[scale=0.6]{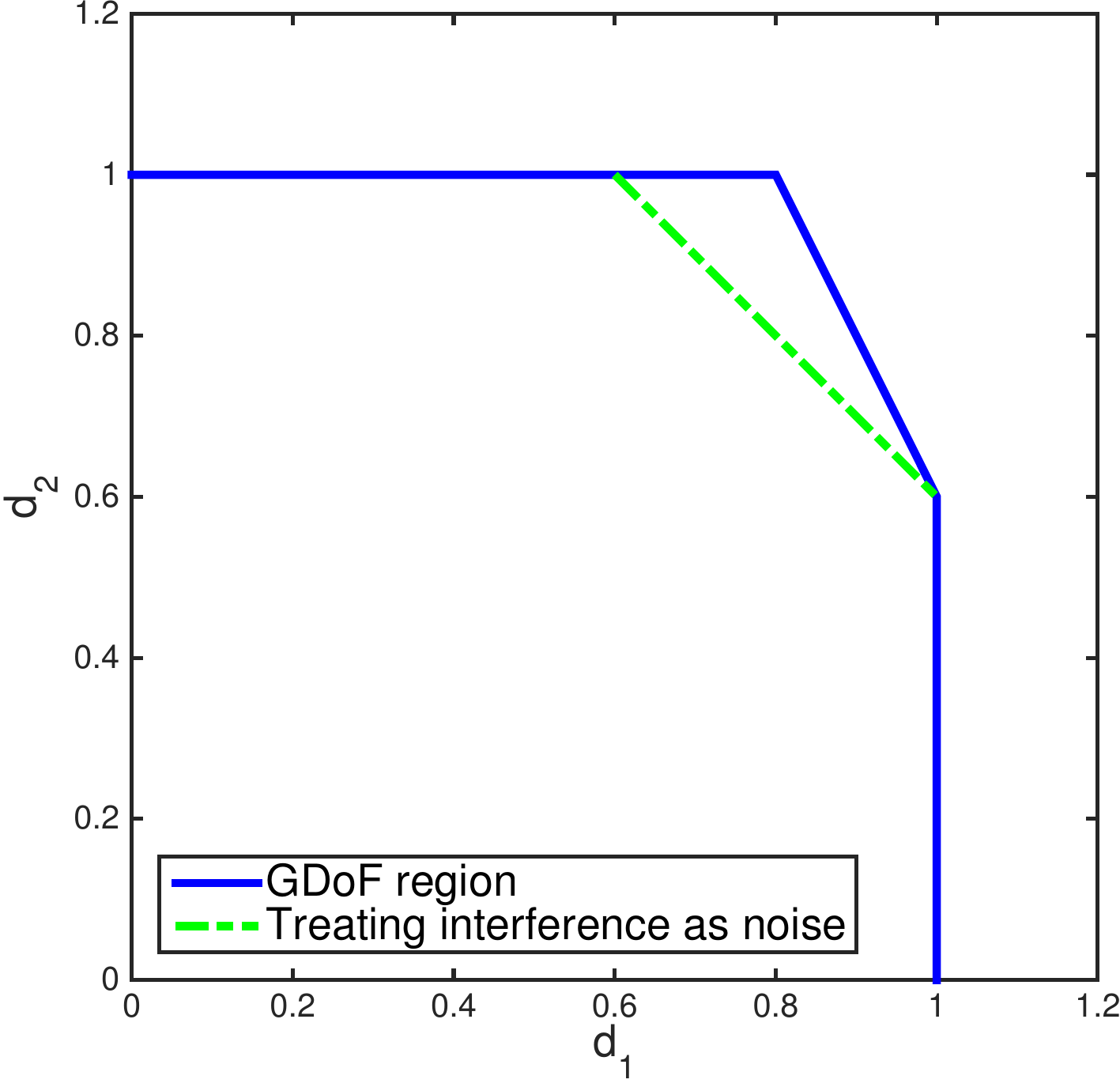}

\caption{\label{fig:TIN}Sub-optimality of treating interference as noise for
the $\left(1,2,1,1\right)$ Z-IC when $\alpha=0.4$.}
\end{figure}

\subsection{Sub-optimality of treating interference as noise}

Our MIMO analysis shows that, unlike the SISO Z-IC with delayed CSIT
and weak interference, treating interference as noise (TIN) at
$R_{1}$ is not in general GDoF-optimal for the MIMO Z-IC with delayed
CSIT, even in the weak interference regime. This is illustrated for
the $\left(1,2,1,1\right)$ MIMO Z-IC with $\alpha=0.4$ in Fig. \ref{fig:TIN},
where the dotted line shows the achievable GDoF region obtained by
treating the interference as noise, which is clearly sub-optimal compared
to the actual GDoF region, shown in the figure with solid lines.

\subsection{Delayed CSIT vs perfect CSIT}

\begin{figure*}[tb]
\subfloat[GDoF region at $\alpha=0.6$.]{\includegraphics[scale=0.32]{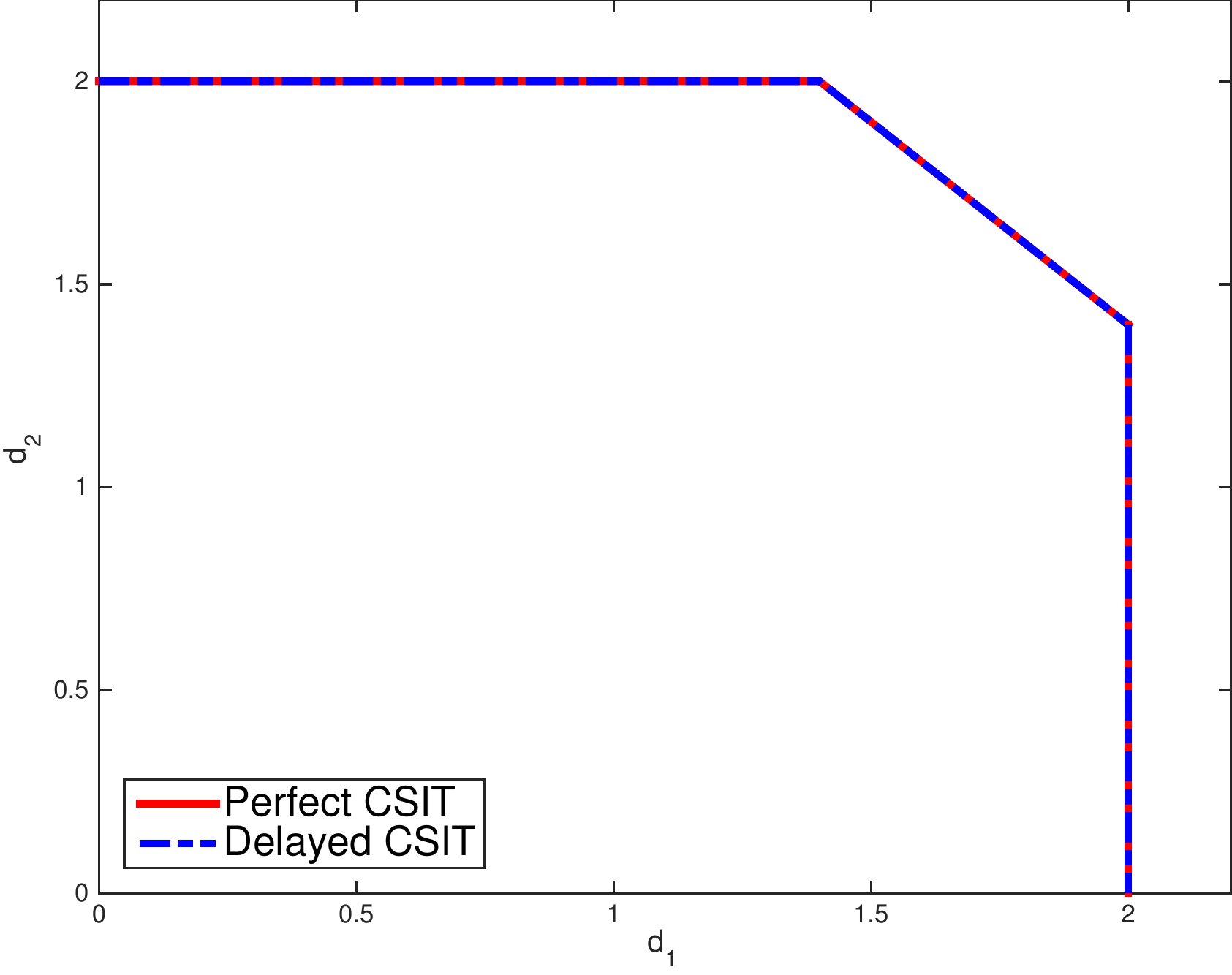}}\hfill
\subfloat[GDoF region at $\alpha=1.4$.]{\includegraphics[scale=0.32]{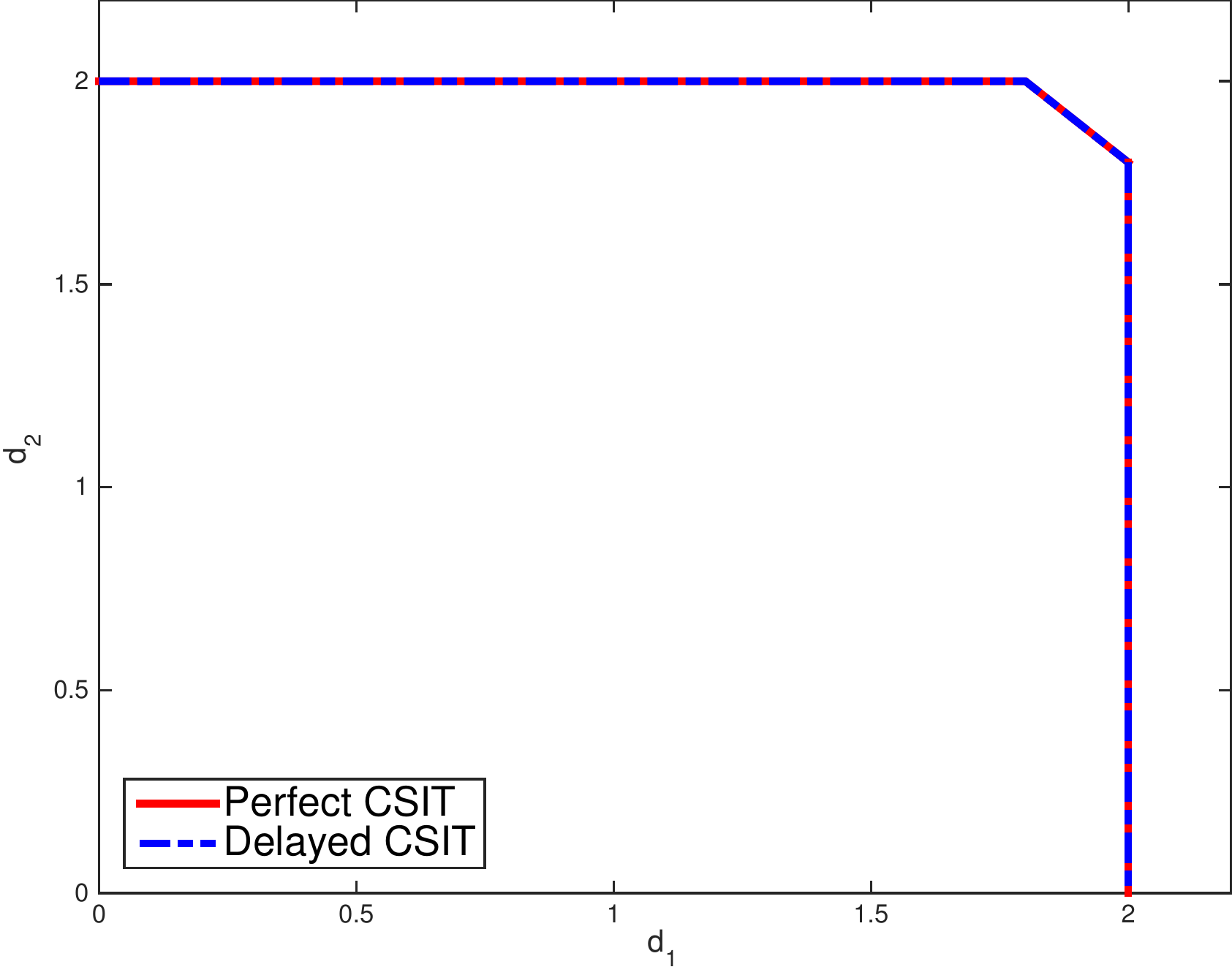}}\hfill
\subfloat[Sum-GDoF at different $\alpha$.]{\includegraphics[scale=0.32]{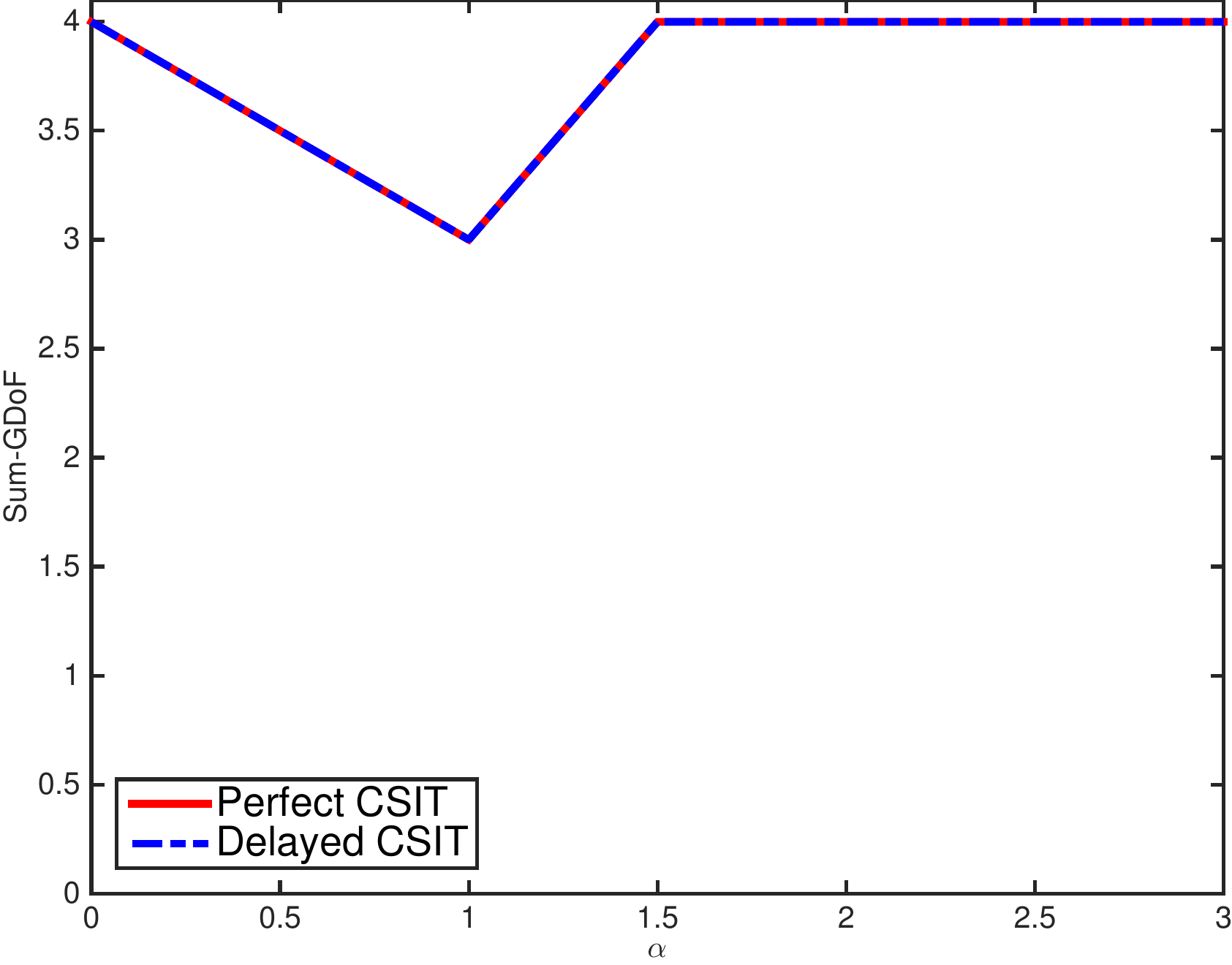}}\hfill

\caption{\label{fig:perf-vs-del-2232}Comparison of the GDoF region and sum-GDoF
of the $\left(2,2,3,2\right)$ Z-IC with delayed and perfect CSIT.}
\end{figure*}
\begin{figure*}[tb]
\subfloat[GDoF region at $\alpha=0.6$.]{\includegraphics[scale=0.32]{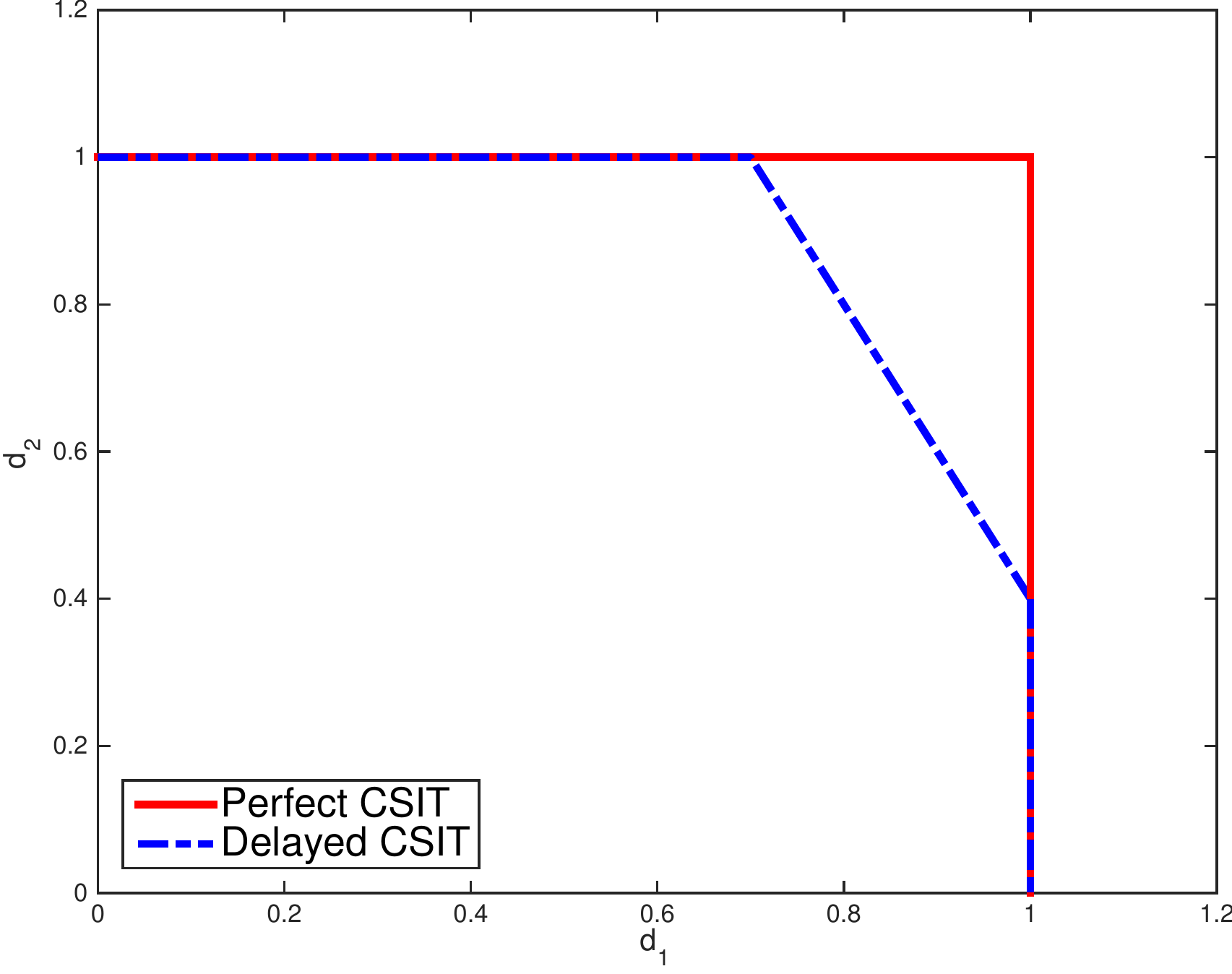}}\hfill
\subfloat[GDoF region at $\alpha=1.4$.]{\includegraphics[scale=0.32]{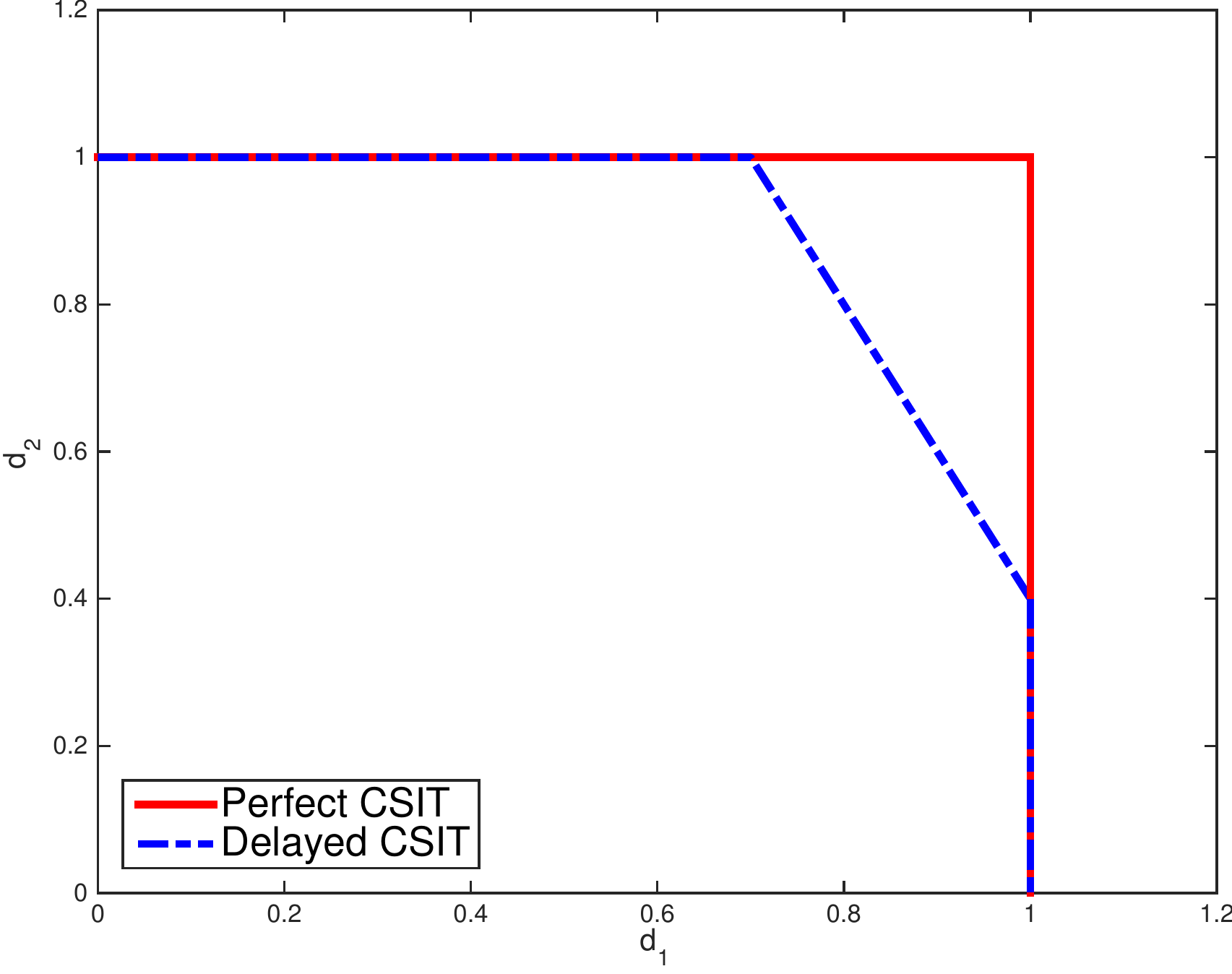}}\hfill
\subfloat[Sum-GDoF at different $\alpha$.]{\includegraphics[scale=0.32]{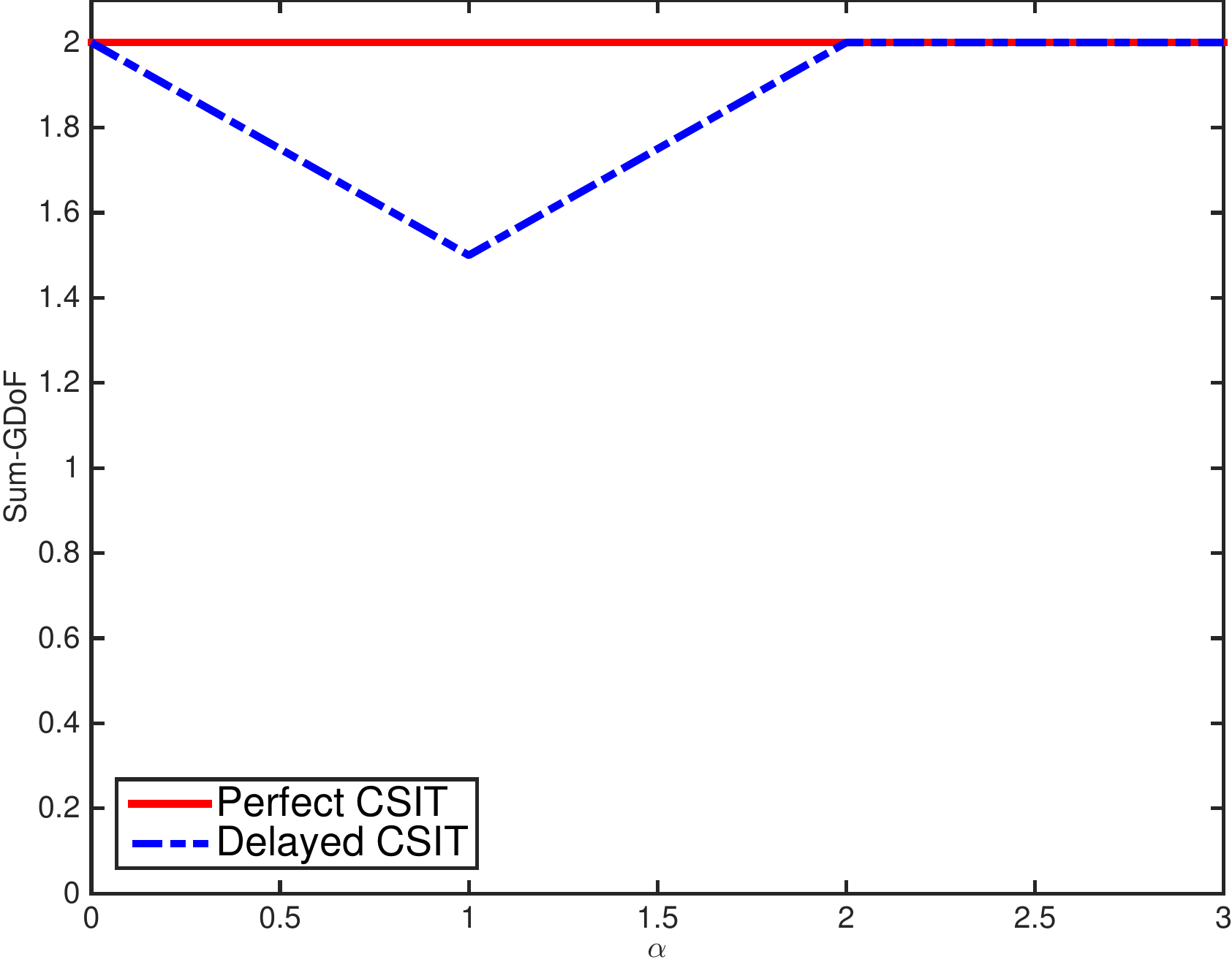}}\hfill

\caption{\label{fig:perf-vs-del-1211}Comparison of the GDoF region and sum-GDoF
of the $\left(1,2,1,1\right)$ Z-IC with delayed CSIT and perfect
CSIT.}
\end{figure*}
\begin{figure*}[tb]
\subfloat[GDoF region at $\alpha=0.6$.]{\includegraphics[scale=0.32]{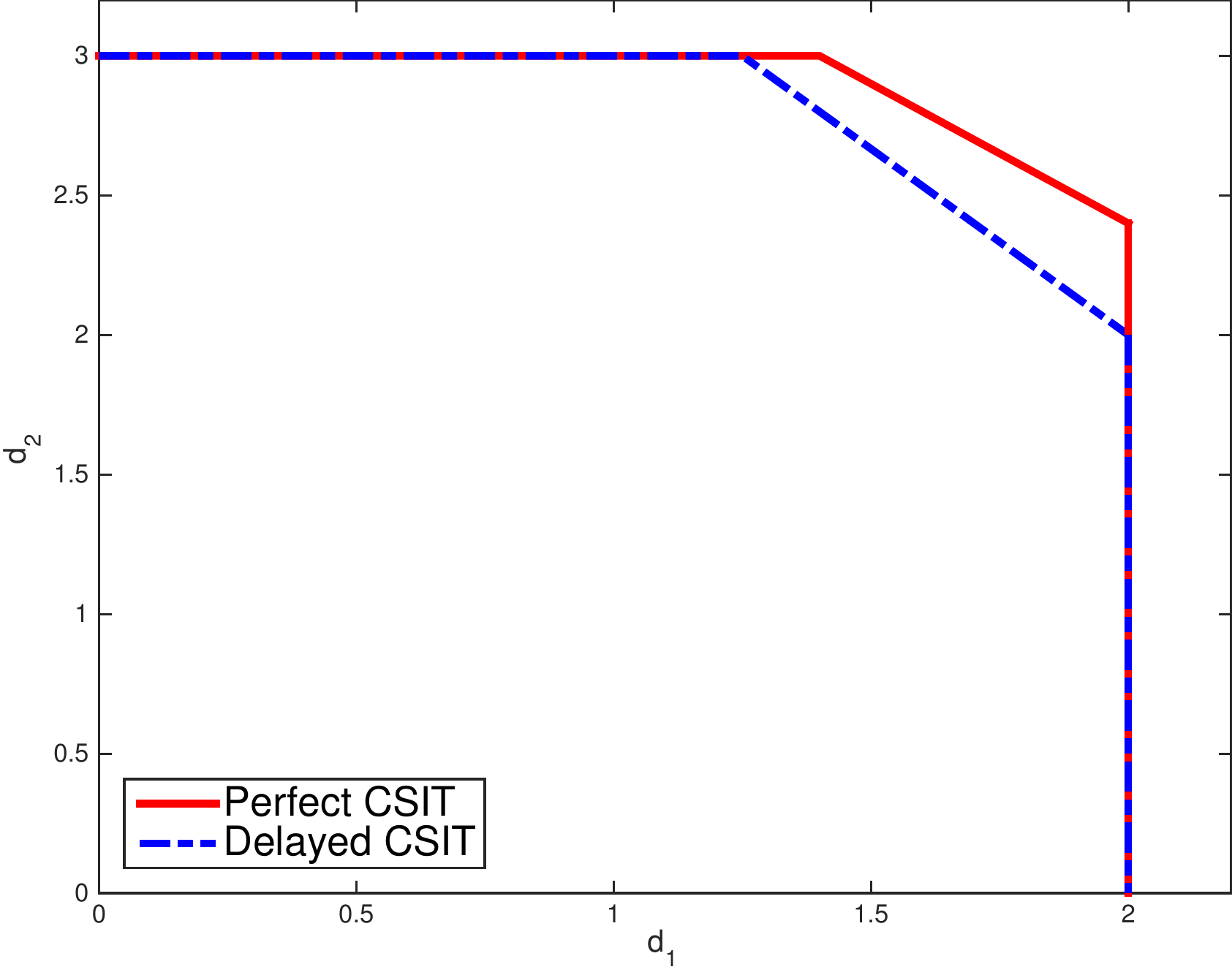}}\hfill
\subfloat[GDoF region at $\alpha=1.4$.]{\includegraphics[scale=0.32]{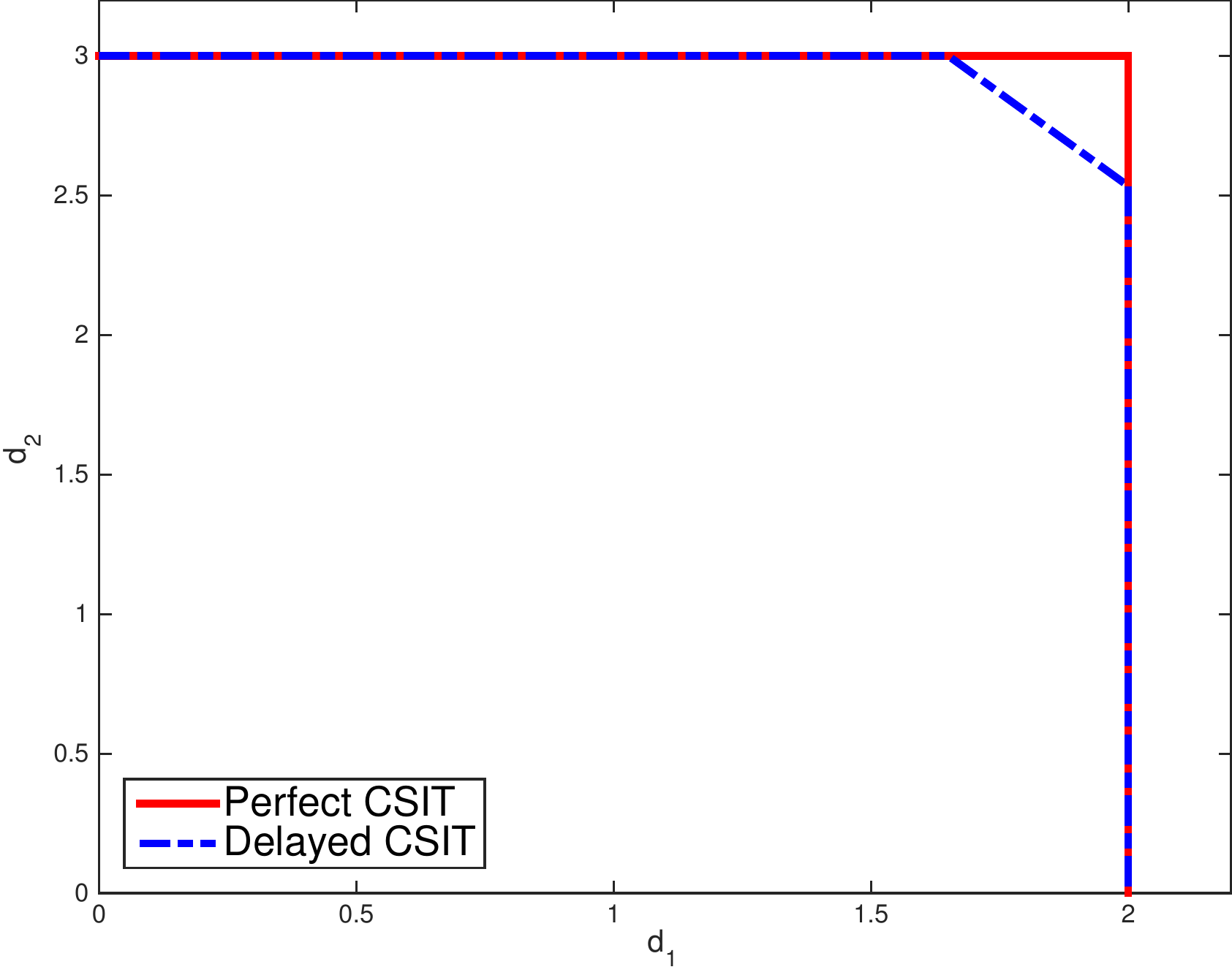}}\hfill
\subfloat[Sum-GDoF at different $\alpha$.]{\includegraphics[scale=0.32]{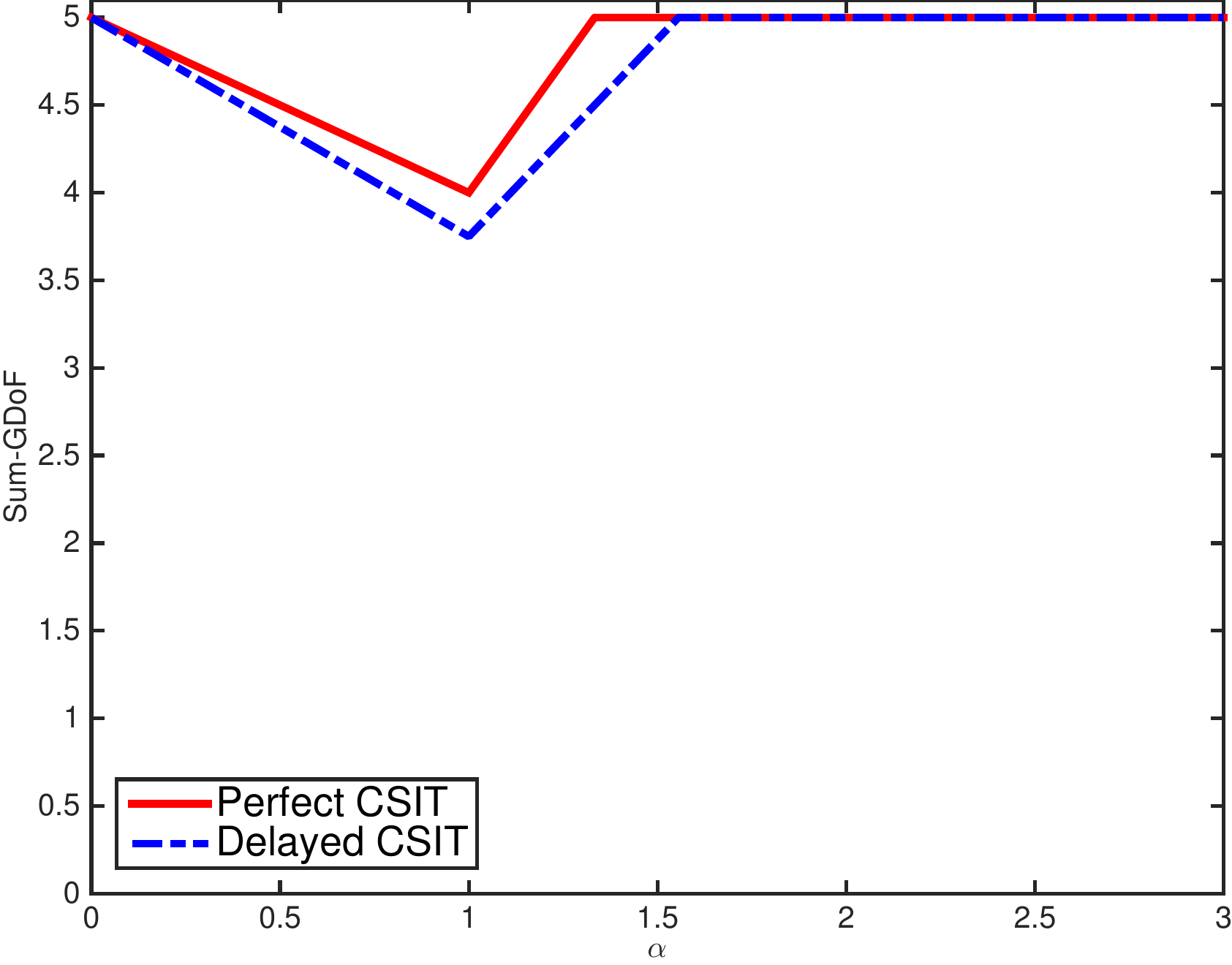}}\hfill

\caption{\label{fig:perf-vs-del-2433}Comparison of the GDoF region and sum-GDoF
of the $\left(2,4,3,3\right)$ Z-IC with delayed CSIT and perfect
CSIT.}
\end{figure*}

In Sections \ref{sec:WEAK-INTERFERENCE} and \ref{sec:STRONG-INTERFERENCE},
we compared the delayed CSIT GDoF region and the corresponding perfect
CSIT GDoF region with weak and strong interference, respectively,
for various antenna configurations. The insights gained from these
comparisons are illustrated below with some representative examples.

When $M_{2}\leq N_{1}$, delayed CSIT is sufficient to achieve the
perfect CSIT GDoF region for all values of $\alpha$. This is illustrated
for the $\left(2,2,3,2\right)$ Z-IC in Fig. \ref{fig:perf-vs-del-2232},
where the delayed CSIT and perfect CSIT GDoF regions are shown to
coincide at $\alpha=0.6$ (weak interference) and $\alpha=1.4$ (strong
interference). Delayed CSIT is also sufficient to achieve the perfect
CSIT GDoF region when $M_{2}>N_{1}$, but only for Case I in Sections
\ref{sec:WEAK-INTERFERENCE}, i.e., $\alpha\leq1$ and $\frac{N_{1}-M_{1}}{N_{1}^{\prime}}\geq\frac{N_{2}}{M_{2}}$,
and \ref{sec:STRONG-INTERFERENCE}, i.e., $\alpha>1$ and \emph{$\alpha\geq1+\frac{N_{2}}{N_{1}^{\prime}}-\frac{M_{2}\left(N_{1}-M_{1}\right)}{\left(N_{1}^{\prime}\right)^{2}}$.}

For the remaining choices of antenna tuples and $\alpha$ when $M_{2}>N_{1}$,
i.e., Case II of Sections \ref{sec:WEAK-INTERFERENCE} and \ref{sec:STRONG-INTERFERENCE},
we have already shown that the delayed CSIT GDoF region is strictly
smaller than the corresponding perfect CSIT GDoF region (excluding
the trivial case of $\alpha=0$). This is illustrated in Figs. \ref{fig:perf-vs-del-1211}
and \ref{fig:perf-vs-del-2433}, for the $\left(1,2,1,1,\right)$
and $\left(2,4,3,3\right)$ Z-IC, respectively. For the $\left(1,2,1,1\right)$
Z-IC, the perfect CSIT sum-GDoF bound in \eqref{eq:perfect-CSIT-GDoF-region}
is never active, and consequently the perfect CSIT GDoF region remains
the same at all values of $\alpha$. For the $\left(2,4,3,3\right)$
Z-IC, both the delayed CSIT and perfect CSIT bounds can be active,
as seen by the shapes of both the GDoF regions at $\alpha=0.6$ and
$\alpha=1.4$.

\subsection{Sum-GDoF}

\begin{figure*}[tb]
\subfloat[\label{fig:1212-weak-int}GDoF region at $\alpha=0.6$.]{\includegraphics[scale=0.32]{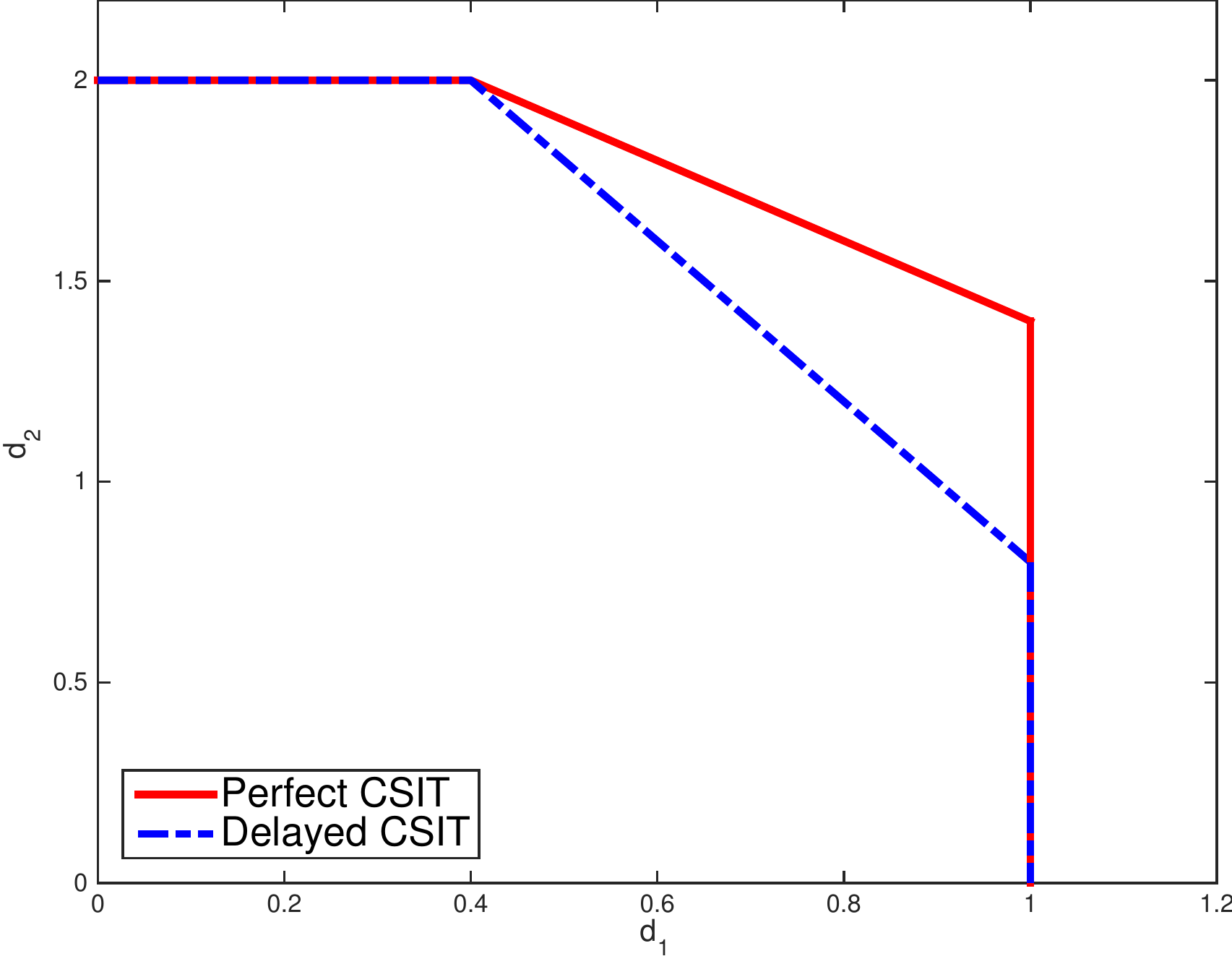}}\hfill
\subfloat[GDoF region at $\alpha=1.4$.]{\includegraphics[scale=0.32]{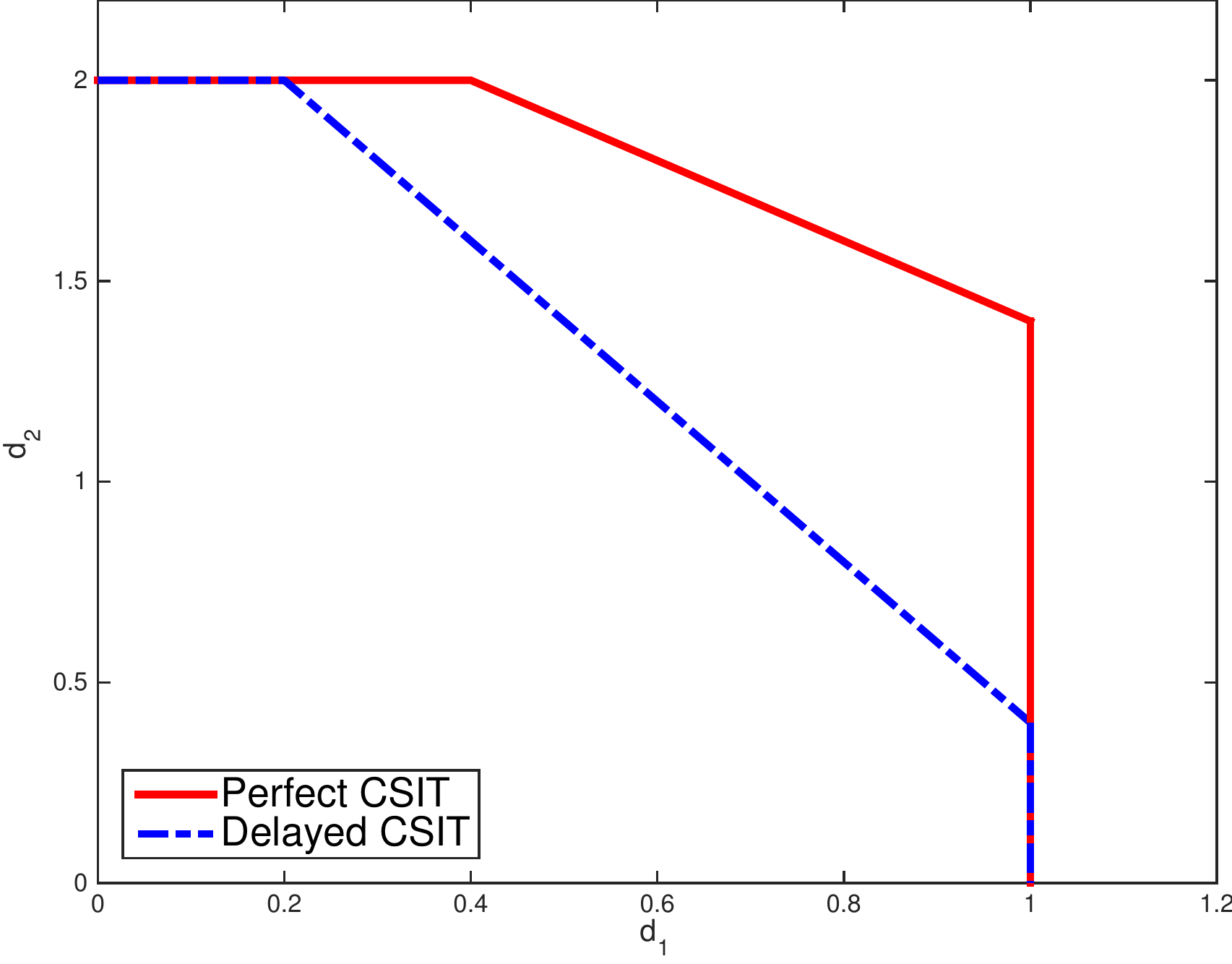}}\hfill
\subfloat[\label{fig:Sum-GDoF-1212}Sum-GDoF at different $\alpha$.]{\includegraphics[scale=0.32]{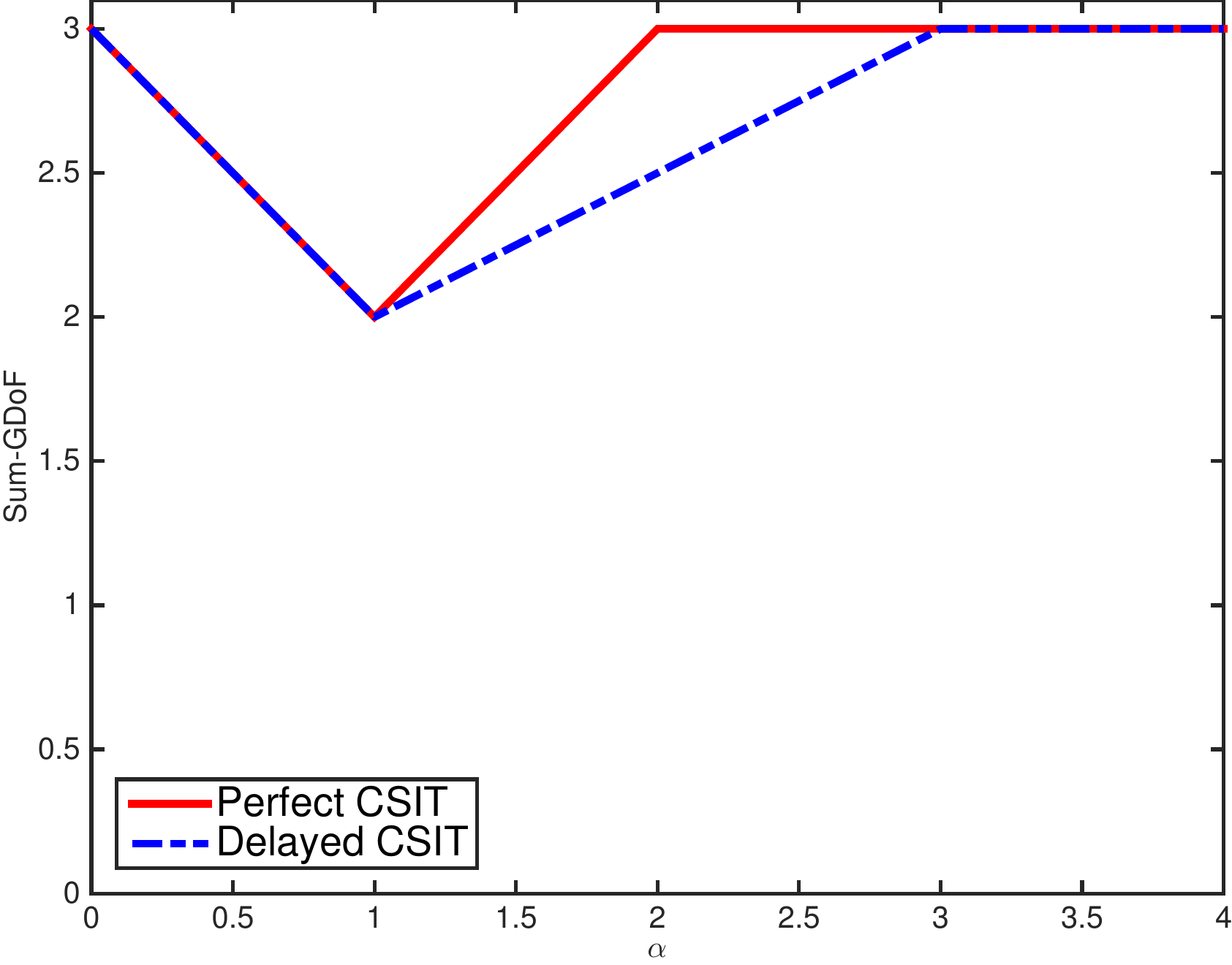}}\hfill

\caption{\label{fig:perf-vs-del-1212}Comparison of the GDoF region and sum-GDoF
of the $\left(1,2,1,2\right)$ Z-IC with delayed CSIT and perfect
CSIT.}
\end{figure*}
In Section \ref{sec:WEAK-INTERFERENCE}, we showed that the sum-GDoF
is a monotonically decreasing linear function of $\alpha$ in the
weak interference regime. For the strong interference regime, it was
shown in Section \ref{sec:STRONG-INTERFERENCE} that the sum-GDoF
is a monotonically increasing linear function of $\alpha$. In general,
this leads to a typical V-shape for the sum-GDoF over the complete
range of $\alpha$. We note that the two segments of this V-shape
can have slopes of different magnitudes. In line with our comparison
of the DoF vs GDoF regions, the minimum $d_{\Sigma}$ is attained
at $\alpha=1$.

From the perfect CSIT sum-GDoF results in Sections \ref{sec:WEAK-INTERFERENCE}
and \ref{sec:STRONG-INTERFERENCE}, it is clear that the perfect CSIT
sum-GDoF is also, in general, a V-shaped function of $\alpha$. We
compare the delayed CSIT sum-GDoF with its perfect CSIT counterpart
in Figs. \ref{fig:perf-vs-del-2232}, \ref{fig:perf-vs-del-1211},
\ref{fig:perf-vs-del-2433} and \ref{fig:perf-vs-del-1212}, for various
antenna configurations. In all of these figures, we see the characteristic
V-shape for the delayed CSIT sum-GDoF, with the minimum occurring
at $\alpha=1$. For the $\left(2,2,3,2\right)$ Z-IC, the GDoF regions
with delayed CSIT and perfect CSIT coincide, and thus, so does the
sum-GDoF, as seen in Fig. \ref{fig:perf-vs-del-2232}. For the $\left(1,2,1,1\right)$
Z-IC, we saw earlier that the perfect CSIT sum-GDoF bound is never
active, and this is seen in the constant value of the perfect CSIT
sum-GDoF over the range of $\alpha$ in Fig. \ref{fig:perf-vs-del-1211}.
The sum-GDoF of the $\left(2,4,3,3\right)$ Z-IC, in Fig. \ref{fig:perf-vs-del-2433},
displays the characteristic V-shape for both the delayed CSIT and
perfect CSIT sum-GDoF curves, albeit with different slopes for the
two segments of the V-shape in both CSIT regimes. It is clear that
for both the $\left(1,2,1,1\right)$ and $\left(2,4,3,3\right)$ Z-IC,
delayed CSIT is not sufficient to attain the perfect CSIT sum-GDoF
in the weak interference regime, and for a range of $\alpha$ in the
strong interference regime.

An interesting phenomenon is observed for the $\left(1,2,1,2\right)$
Z-IC, as seen in Fig. \ref{fig:perf-vs-del-1212}. While delayed CSIT
is insufficient in this case to achieve the complete perfect CSIT
GDoF region in the weak interference regime (except for when $\alpha=0$),
the sum-GDoF plot in Fig. \ref{fig:Sum-GDoF-1212} shows that, for
weak interference, delayed CSIT is still sufficient to achieve the
perfect CSIT sum-GDoF for the $\left(1,2,1,2\right)$ Z-IC. The clue
to this behavior can be found in Fig. \ref{fig:1212-weak-int}, where
we see that, for weak interference ($\alpha=0.6$ in this case), the
perfect CSIT and delayed CSIT GDoF regions with weak interference
share the GDoF corner-point (point $P_{1}$ in Fig. \ref{fig:GDoF-region-Case-II})
at which the maximum sum-GDoF is achieved. By analyzing the corner
points of the delayed CSIT and perfect CSIT GDoF regions in the weak
interference regime to see where they coincide, it is easy to show
that this phenomenon, where delayed CSIT is sufficient to achieve
the perfect CSIT sum-GDoF with weak interference in spite of the perfect
CSIT GDoF region being strictly larger than its delayed CSIT counterpart,
occurs in Case II of Section \ref{sec:WEAK-INTERFERENCE} when 
\[
N_{2}=M_{2}>N_{1}.
\]
Fig. \ref{fig:perf-vs-del-1212} also shows that, for the $\left(1,2,1,2\right)$
Z-IC, delayed CSIT still remains insufficient to achieve the perfect
CSIT sum-GDoF in the strong interference regime (when the delayed
CSIT bound is active).

\section{CONCLUSION\label{sec:CONCLUSION}}

In this paper, we characterize the GDoF region of the $\left(M_{1},M_{2},N_{1},N_{2}\right)$
MIMO Z-IC under the assumption of delayed CSIT. We obtain a new outer
bound by using a combination of a genie, the extremal inequality and
maximizing the weighted sum-rate of the two users in the high SNR
regime. We next develop a general block-Markov achievability scheme
that uses interference quantization to take advantage of the different
power levels of the INR and SNR. By specializing this achievability
scheme to both weak and strong interference regimes, we show that
the outer bound region in each regime is achievable. The GDoF region
is found to be equal to or larger than the DoF region over the whole
range of $\alpha$ and for all antenna tuples. Moreover, the antenna
configurations for which delayed CSIT is sufficient to achieve the
perfect CSIT GDoF region or sum-GDoF are characterized. Even in the
weak interference regime, we show that treating interference as noise
is not a GDoF-optimal strategy in general. To the best of our knowledge,
this is the first paper to characterize of the GDoF region of any
network with distributed transmitters and delayed CSIT, as well as
the first paper to characterize the GDoF region of a MIMO network
with delayed CSIT and arbitrary number of antennas at each node. An
investigation into the effect of disparate link strengths on the capacity
of other MIMO networks, e.g., the 2-user IC, under channel uncertainty,
and delayed CSIT in particular, through such GDoF characterization
remains an interesting avenue for future research.

\appendices{}

\section{Proof of Lemma \ref{lem:rank-optimization-lemma}\label{sec:Outer-Bound-Optimization}}

To prove Lemma \ref{lem:rank-optimization-lemma}, we show that the
function%
{} $g\left(r\right)$ defined in \eqref{eq:rank-opt-expression-to-optimize}
is maximized when $r=0$. Without any loss of generality, we assume
that $M_{2}\leq N_{1}+N_{2}$, since it is clear that the function
$g\left(r\right)$ remains unchanged irrespective of whether $M_{2}-r>N_{1}+N_{2}$
or $M_{2}-r=N_{1}+N_{2}$. Below, we analyze the strong interference
and weak interference cases separately.

\subsection*{Strong interference}

We analyze the function $g\left(r\right)$ in the strong interference
regime $\left(\alpha>1\right)$ on a case-by-case basis, as follows:

\emph{Case 1)} when $M_{2}\leq N_{1}$: %
$g\left(r\right)$ can be written as 
\[
\frac{f\left(N_{1},\left(1,M_{1}\right),\left(\alpha,M_{2}-r\right)\right)}{M_{2}},
\]
 which is a monotonically decreasing function of $r$, and is thus
maximized at $r=0$. 

\emph{Case 2)} when $M_{2}>N_{1}$: the first term in $g\left(r\right)$
can be written as 
\begin{eqnarray}
\frac{f\left(N_{1},\left(1,M_{1}\right),\left(\alpha,M_{2}-r\right)\right)}{\min\left(M_{2},N_{1}\right)}\qquad\qquad\qquad\qquad\nonumber \\
\negmedspace=\negmedspace\frac{\alpha\min\negmedspace\left(M_{2}\negmedspace-\negmedspace r,N_{1}\right)\negmedspace+\negmedspace\min\negmedspace\left(\left(N_{1}\negmedspace-\negmedspace\left(M_{2}\negmedspace-\negmedspace r\right)\right)^{+}\negmedspace,M_{1}\right)}{N_{1}}.\label{eq:rank-opt-strong-int-substitute}
\end{eqnarray}
Hence, $g\left(r\right)$ becomes 
\begin{equation}
\frac{\min\left(\left(N_{1}\negthinspace-\negthinspace\left(M_{2}-r\right)\right)^{+},M_{1}\right)}{N_{1}}\negmedspace+\negmedspace\frac{f\left(M_{2}\negmedspace-\negmedspace r,\left(\alpha,N_{1}\right),\left(1,N_{2}\right)\right)}{M_{2}}.\label{eq:rank-opt-strong-int-M2-greater-N1}
\end{equation}
The first term in \eqref{eq:rank-opt-strong-int-M2-greater-N1} is
monotonically increasing with $r$, while the second term decreases
monotonically with $r$. 

Compared to the full-rank $\left(r=0\right)$ case, it can be shown
that any increase in the first term, from setting $r>0$, is always
offset by a corresponding decrease in the second term. We illustrate
this by analyzing the most optimistic scenario, where $g\left(r\right)$
attains its maximum value for $r>0$, i.e., when $r=M_{2}$ and $M_{1}\geq N_{1}$.
In this scenario, the first term increases by $\frac{N}{N}=1$ over
the full-rank case, while the corresponding loss in the second term
is 
\[
\frac{\alpha N_{1}+M_{2}-N_{1}}{M_{2}}=1+\left(\alpha-1\right)\frac{N_{1}}{M_{2}}.
\]
Thus, the maximum possible increase in $g\left(r\right)$, for $r>0$,
compared to the full-rank case is 
\[
1-\left[1+\left(\alpha-1\right)\frac{N_{1}}{M_{2}}\right]=-\left(\alpha-1\right)\frac{N_{1}}{M_{2}}<0,
\]
proving that $g\left(r\right)$ is maximized at $r=0$.

\subsection*{Weak interference}

Since $\alpha\leq1$, the second term of $g\left(r\right)$ can be
written as:%
\begin{equation}
\frac{f\negthinspace\left(M_{2}\negmedspace-\negmedspace r,\left(\alpha,N_{1}\negthinspace\right)\negthinspace,\left(1,N_{2}\negthinspace\right)\right)}{\min\left(M_{2},N_{1}\negmedspace+\negmedspace N_{2}\right)}\negthinspace=\negthinspace\frac{f\negthinspace\left(M_{2},\left(\alpha,N_{1}\right),\left(1,N_{2}\right)\right)}{\min\left(M_{2},N_{1}+N_{2}\right)}-l_{2}\left(r\right)\negthinspace,\label{eq:l2}
\end{equation}
where the loss function $l_{2}\left(r\right)$, depends on the antenna
tuple as follows (recall that $M_{2}\leq N_{1}+N_{2}$):\emph{}\\
\emph{Case 1)} when $M_{2}\leq N_{2}$: 
\[
l_{2}\left(r\right)=\frac{r}{M_{2}}.
\]
\emph{Case 2)} when $M_{2}>N_{2}$:%
{} 
\[
l_{2}\left(r\right)=\begin{cases}
\frac{r\alpha}{M_{2}}, & r\leq M_{2}-N_{2}\\
\frac{\left(M_{2}-N_{2}\right)\alpha+\left(N_{2}-\left(M_{2}-r\right)\right)}{M_{2}}, & r>M_{2}-N_{2}.
\end{cases}
\]
The third term of $g\left(r\right)$ can also be written as  
\begin{eqnarray}
\frac{\alpha\min\left(M_{2}-r,N_{1}\right)}{\min\left(M_{2},N_{1}\right)} & = & \alpha-l_{3}\left(r\right),\label{eq:l3}
\end{eqnarray}
where the loss function $l_{3}\left(r\right)$ depends on the antenna
tuple as follows:\emph{}\\
\emph{Case 1)} when $M_{2}\leq N_{1}$: 
\[
l_{3}\left(r\right)=\frac{\alpha r}{M_{2}}.
\]
\emph{Case 2)} when $M_{2}>N_{1}$: 
\[
l_{3}\left(r\right)=\begin{cases}
0, & r\leq M_{2}-N_{1}\\
\frac{N_{1}-\left(M_{2}-r\right)}{N_{1}}\alpha, & r>M_{2}-N_{1}.
\end{cases}
\]
Now, by substituting \eqref{eq:l2} and \eqref{eq:l3}, we can write
$g\left(r\right)$ as 
\begin{eqnarray*}
x\left(r\right)+\left[l_{3}\left(r\right)-l_{2}\left(r\right)\right],
\end{eqnarray*}
where 
\[
x\left(r\right)\negthinspace=\negthinspace\frac{f\negthinspace\left(N_{1},\negthinspace\left(1,M_{1}\right)\negthinspace,\negthinspace\left(\alpha,M_{2}\negthinspace-\negthinspace r\right)\negthinspace\right)\negthinspace}{N_{1}}+\frac{f\negthinspace\left(M_{2},\left(\alpha,N_{1}\right)\negthinspace,\negthinspace\left(1,\negthinspace N_{2}\right)\negthinspace\right)\negthinspace}{M_{2}}-\alpha
\]
is a monotonically decreasing function of $r$. Thus, to prove that
the function $g\left(r\right)$ is a monotonically decreasing function
of $r$, which consequently attains it maximum at $r=0$, it suffices
to prove that 
\begin{equation}
l_{3}\left(r\right)-l_{2}\left(r\right)\leq0,\label{eq:l3-l2}
\end{equation}
for $0\leq r\leq M_{2}$. We prove \eqref{eq:l3-l2} below on a case-by-case
basis, for all antenna tuples.\emph{}\\
\emph{Case 1)} when $M_{2}\leq N_{1}$, $M_{2}\leq N_{2}$: 
\[
l_{3}\left(r\right)-l_{2}\left(r\right)=\left(\alpha-1\right)\frac{r}{M_{2}}\leq0.
\]
\emph{Case 2)} when $N_{2}<M_{2}\leq N_{1}$: 
\begin{eqnarray*}
l_{3}\left(r\right)-l_{2}\left(r\right) & = & \begin{cases}
\left(\alpha-1\right)\frac{r}{M_{2}}, & r\leq M_{2}-N_{2}\\
\left(\alpha-1\right)\frac{N_{2}-\left(M_{2}-r\right)}{M_{2}}, & r>M_{2}-N_{2}
\end{cases}\\
 & \leq & 0.
\end{eqnarray*}
\emph{Case 3)} when $N_{1}<M_{2}\leq N_{2}$: In this case, there
are two sub-cases, which we analyze separately below.

When $r\leq M_{2}-N_{1}$, we have
\[
l_{3}\left(r\right)-l_{2}\left(r\right)=0-\frac{r}{M_{2}}\leq0.
\]
When $r>M_{2}-N_{1}$, we have 
\begin{eqnarray*}
l_{3}\left(r\right)-l_{2}\left(r\right) & = & \frac{\left(N_{1}-M_{2}+r\right)\alpha}{N_{1}}-\frac{r}{M_{2}}\\
 & = & \left(1-\frac{M_{2}}{N_{1}}\right)\alpha+r\left(\frac{\alpha}{N_{1}}-\frac{1}{M_{2}}\right)\\
 & \overset{\left(a\right)}{\leq} & \left(1-\frac{M_{2}}{N_{1}}\right)\alpha+M_{2}\left(\frac{\alpha}{N_{1}}-\frac{1}{M_{2}}\right)^{+}\\
 & \overset{\left(b\right)}{\leq} & 0,
\end{eqnarray*}
where we have used $r\leq M_{2}$ in $\left(a\right)$ and $N_{1}<M_{2}$
in $\left(b\right)$.\emph{}\\
\emph{Case 4)} when $N_{1}<M_{2}$, $N_{2}<M_{2}$: We have three
sub-cases, which are analyzed separately below. 

When $r\leq M_{2}-N_{1}$, we have 
\[
l_{3}\left(r\right)-l_{2}\left(r\right)=0-l_{2}\left(r\right)\leq0.
\]
The second sub-case is defined as $M_{2}-N_{1}<r\leq M_{2}-N_{2}$,
and we obtain
\begin{eqnarray*}
l_{3}\left(r\right)-l_{2}\left(r\right) & = & \frac{\left(N_{1}-M_{2}+r\right)\alpha}{N_{1}}-\frac{r\alpha}{M_{2}}\\
 & = & \left(1-\frac{M_{2}}{N_{1}}\right)\left(1-\frac{r}{M_{2}}\right)\alpha\\
 & \leq & 0.
\end{eqnarray*}
Thirdly, when $r>M_{2}-N_{1}$ and $r>M_{2}-N_{2}$, we have 
\begin{eqnarray*}
 &  & \quad l_{3}\left(r\right)-l_{2}\left(r\right)\\
 &  & =\frac{\left(N_{1}-M_{2}+r\right)\alpha}{N_{1}}-\frac{\left(M_{2}-N_{2}\right)\alpha+\left(N_{2}-M_{2}+r\right)}{M_{2}}\\
 &  & =\left(\frac{M_{2}-N_{2}-r}{M_{2}}\right)-\alpha\left(\frac{M_{2}-r}{N_{1}}-\frac{N_{2}}{M_{2}}\right)\\
 &  & \overset{\left(a\right)}{\leq}\left(\frac{M_{2}-N_{2}-r}{M_{2}}\right)-\alpha\left(\frac{M_{2}-r}{M_{2}}-\frac{N_{2}}{M_{2}}\right)\\
 &  & =\left(\frac{M_{2}-N_{2}-r}{M_{2}}\right)\left(1-\alpha\right)\\
 &  & \overset{\left(b\right)}{\leq}0,
\end{eqnarray*}
where we have used $N_{1}<M_{2}$ in $\left(a\right)$ and $M_{2}-N_{2}-r<0$
in $\left(b\right)$. %

Thus, for all antenna tuples, we have shown that the function $g\left(r\right)$
is a monotonically decreasing function of $r$ in the weak interference
regime, and is hence maximized at $r=0$.

\section{Achievable GDoF Region\label{sec:Achievable-GDoF-Region}}

In this appendix, we show that the achievable GDoF region for the
MIMO Z-IC using the general achievability scheme from Section \ref{sec:ACHIEVABLE-SCHEME}
is given by \eqref{eq:gdof-zic-equation-range-start}-\eqref{eq:gdof-zic-equation-d2-deta}.
The 2-user MACs obtained in \eqref{eq:R1-MAC} and \eqref{eq:R2-MAC}
are shown below (we omit the block index $b$ for the channel matrices):
\begin{align*}
\underbrace{y_{1}\left[b\right]-\eta_{b}}_{Y_{1}^{\prime}} & =\sqrt{\rho^{\alpha}}H_{12}\underbrace{x_{2c}\left(l_{b-1}\right)}_{X_{c}^{\prime}}+\sqrt{\rho}H_{11}\underbrace{u_{1}\left(w_{1,b}\right)}_{X_{1}^{\prime}}+Z_{1},
\end{align*}
\[
\underbrace{\left[\begin{array}{c}
y_{2}\left[b\right]\\
\eta_{b}
\end{array}\right]}_{Y_{2}^{\prime}}\negmedspace=\negmedspace\left[\begin{array}{c}
\sqrt{\rho}H_{22}\negmedspace\\
0
\end{array}\right]\underbrace{\negmedspace x_{2c}\negmedspace\left(l_{b-1}\right)\negmedspace}_{X_{c}^{\prime}}+\negmedspace\left[\begin{array}{c}
\negmedspace\sqrt{\rho}H_{22}\negmedspace\negmedspace\\
\negmedspace\sqrt{\rho^{\alpha}}H_{12}\negmedspace\negmedspace
\end{array}\right]\underbrace{\negmedspace u_{2}\negmedspace\left(w_{2,b}\right)}_{X_{2}^{\prime}}+Z_{2},
\]
where, $Z_{i}$ is the AWGN at receiver $R_{i}$, and, for simplicity,
we have also defined $Y_{1}^{\prime}$, $Y_{2}^{\prime}$, $X_{c}^{\prime}$,
$X_{1}^{\prime}$ and $X_{2}^{\prime}$, such that $X_{c}^{\prime}$,
$X_{1}^{\prime}$ and $X_{2}^{\prime}$ have achievable rates $\bar{R}_{c}$,
$\bar{R}_{1}$ and $\bar{R}_{2}$, respectively, and the associated
GDoF are respectively $d_{\eta}$, $d_{1b}$ and $d_{2b}$ (from \eqref{eq:gdof-zic-equation-range-start}-\eqref{eq:gdof-zic-equation-d2-deta}).
We note that $X_{c}^{\prime}\sim\mathcal{CN}\left(0,Q_{c}\right)$,
$X_{1}^{\prime}\sim\mathcal{CN}\left(0,Q_{1}\right)$ and $X_{2}^{\prime}\sim\mathcal{CN}\left(0,Q_{2}\right)$
are independent, with covariance matrices $Q_{c}\triangleq\mathbf{I}_{M_{2}}$,
$Q_{1}\triangleq\mathbf{I}_{M_{1}}$ and $Q_{2}\triangleq\rho^{-A_{2}}\mathbf{I}_{M_{2}}$.
From \cite{Gamal2012}, the achievable rate region for the 2-user
MAC at receiver $R_{i}$, $i\in\left\{ 1,2\right\} $ is given by
the following information-theoretic inequalities :
\begin{eqnarray}
\bar{R}_{c} & \leq & I\left(X_{c}^{\prime};Y_{i}^{\prime}\left|X_{i}^{\prime},\mathcal{H}^{n}\right.\right),\label{eq:info-theory-rc}\\
\bar{R}_{i} & \leq & I\left(X_{i}^{\prime};Y_{i}^{\prime}\left|X_{c}^{\prime},\mathcal{H}^{n}\right.\right),\label{eq:info-theory-ri}\\
\bar{R}_{c}+\bar{R}_{i} & \leq & I\left(X_{c}^{\prime},X_{i}^{\prime};Y_{i}^{\prime}\left|\mathcal{H}^{n}\right.\right).\label{eq:info-theory-rc-ri}
\end{eqnarray}
From \eqref{eq:info-theory-rc}, we obtain the following GDoF bounds:
\begin{eqnarray}
\bar{R}_{c} & \leq & I\left(X_{c}^{\prime};Y_{1}^{\prime}\left|X_{1}^{\prime},\mathcal{H}^{n}\right.\right)\nonumber \\
 & = & h\left(Y_{1}^{\prime}\left|X_{1}^{\prime},\mathcal{H}^{n}\right.\right)-h\left(Y_{1}^{\prime}\left|X_{1}^{\prime},X_{c}^{\prime},\mathcal{H}^{n}\right.\right)\nonumber \\
 & = & \log\left|\mathbf{I}_{N_{1}}+\rho^{\alpha}H_{12}Q_{c}H_{12}^{\dagger}\right|+\mathcal{O}\left(1\right)\nonumber \\
 & \overset{\left(a\right)}{=} & \alpha\min\left(M_{2},N_{1}\right)\log\rho+O\left(1\right)\nonumber \\
\Rightarrow d_{\eta} & \leq & \alpha N_{1}^{\prime},\label{eq:appendix-deta-intermed-1}
\end{eqnarray}
where we have substituted $Q_{c}=\mathbf{I}_{M_{2}}$, $\left(a\right)$
follows from Lemma \ref{lem:MAC-approx}, and finally dividing both
sides by $\log\rho$ as $\rho\rightarrow\infty$ gives the GDoF bound
\eqref{eq:appendix-deta-intermed-1}. Similarly, we obtain 
\begin{eqnarray}
\bar{R}_{c} & \leq & I\left(X_{c}^{\prime};Y_{2}^{\prime}\left|X_{2}^{\prime},\mathcal{H}^{n}\right.\right)\nonumber \\
 & = & \log\left|\mathbf{I}_{N_{2}}+H_{22}Q_{c}H_{22}^{\dagger}\right|+\mathcal{O}\left(1\right)\nonumber \\
 & \overset{\left(a\right)}{=} & \min\left(M_{2},N_{2}\right)\log\rho+O\left(1\right)\nonumber \\
\Rightarrow d_{\eta} & \overset{\left(b\right)}{\leq} & N_{2},\label{eq:appendix-deta-intermed-2}
\end{eqnarray}
where in $\left(a\right)$, we have substituted $Q_{c}=\mathbf{I}_{M_{2}}$
and used Lemma \ref{lem:MAC-approx}, and $\left(b\right)$ uses the
antenna assumption $N_{2}\leq M_{2}$ from \eqref{eq:gdof-zic-antenna-assumptions-end}.
Combining \eqref{eq:appendix-deta-intermed-1} and \eqref{eq:appendix-deta-intermed-2},
we obtain \eqref{eq:gdof-zic-equation-range-start}.

The bound \eqref{eq:gdof-zic-equation-d1} for $d_{1b}$ follows from
\eqref{eq:info-theory-ri}, with $i=1$, in a similarly straightforward
manner. Bound \eqref{eq:gdof-zic-equation-d1-deta} is obtained from
\eqref{eq:info-theory-rc-ri}, with $i=1$, by a direct application
of Lemma \ref{lem:MAC-approx}, as shown below:\vspace{-4pt}
\begin{eqnarray*}
 &  & \bar{R}_{c}+\bar{R}_{1}\\
 & \leq & I\left(X_{c}^{\prime},X_{1}^{\prime};Y_{1}^{\prime}\right)\\
 & = & \log\left|\mathbf{I}_{N_{1}}+\rho^{\alpha}H_{12}Q_{c}H_{12}^{\dagger}+\rho H_{11}Q_{1}H_{11}^{\dagger}\right|+\mathcal{O}\left(1\right)\\
 & = & f\left(N_{1},\left(\alpha,M_{2}\right),\left(1,M_{1}\right)\right)\log\rho+\mathcal{O}\left(1\right),
\end{eqnarray*}
where we have substituted $Q_{c}=\mathbf{I}_{M_{2}}$ and $Q_{1}=\mathbf{I}_{M_{1}}$,
and dividing both sides by $\log\rho$ gives the required bound \eqref{eq:gdof-zic-equation-d1-deta}.

To prove \eqref{eq:gdof-zic-equation-d2}, we use the singular value
decomposition (SVD) of $\left[\negmedspace\begin{array}{c}
H_{22}\\
H_{12}
\end{array}\negmedspace\right]\negmedspace=\negmedspace U\Lambda V^{\dagger}$, where $U\negmedspace\in\negmedspace\mathbb{C}^{\left(N_{1}+N_{2}\right)\times M_{2}}$
is a matrix such that $U^{\dagger}U=\mathbf{I}_{M_{2}}$, $\Lambda$
is a $M_{2}\times M_{2}$ diagonal matrix such that the diagonal elements
consist of the singular values of $\left[\begin{array}{c}
H_{22}\\
H_{12}
\end{array}\right]$ and $V\in\mathbb{C}^{M_{2}\times M_{2}}$ is matrix such that $V^{\dagger}V=\mathbf{I}_{M_{2}}$.
Using \eqref{eq:info-theory-ri}, we obtain \eqref{eq:gdof-zic-equation-d2}
as follows (we have omitted the $\mathcal{O}\left(1\right)$ terms):\vspace{-4pt}
\begin{eqnarray*}
 &  & \bar{R}_{2}\leq I\left(X_{2}^{\prime};Y_{2}^{\prime}\left|X_{c}^{\prime},\mathcal{H}^{n}\right.\right)\\
 &  & \negthickspace\negthickspace\negthickspace=\log\left|\mathbf{I}_{N_{2}+N_{1}\negthickspace}+\negthickspace\left[\begin{array}{c}
\sqrt{\rho}H_{22}\\
\sqrt{\rho^{\alpha}}H_{12}
\end{array}\right]Q_{2}\left[\begin{array}{cc}
\sqrt{\rho}H_{22}^{\dagger} & \sqrt{\rho^{\alpha}}H_{12}^{\dagger}\end{array}\right]\right|\\
 &  & \negthickspace\negthickspace\negthickspace\overset{\left(a\right)}{=}\log\left|\mathbf{I}\negthickspace+\negthickspace\left[\negthickspace\begin{array}{cc}
\rho^{\left(1-A_{2}\right)}I_{N_{2}}\negthickspace\negthickspace & 0\\
0\negthickspace\negthickspace & \rho^{\left(\alpha-A_{2}\right)}I_{N_{1}}
\end{array}\negthickspace\right]\negthickspace\left[\negthickspace\begin{array}{c}
H_{22}\\
H_{12}
\end{array}\negthickspace\right]\negthickspace\left[\negthickspace\begin{array}{cc}
H_{22}^{\dagger}\negthickspace\negthickspace & H_{12}^{\dagger}\negthickspace\end{array}\right]\right|\\
 &  & \negthickspace\negthickspace\negthickspace\overset{\left(b\right)}{=}\log\left|\mathbf{I}+\left[\negthickspace\begin{array}{cc}
\rho^{\left(1-A_{2}\right)}I_{N_{2}}\negthickspace\negthickspace & 0\\
0\negthickspace\negthickspace & \rho^{\left(\alpha-A_{2}\right)}I_{N_{1}}
\end{array}\negthickspace\right]U\Lambda\underbrace{V^{\dagger}V}_{\mathbf{I}_{M_{2}}}\Lambda U^{\dagger}\right|\\
 &  & \negthickspace\negthickspace\negthickspace\overset{\left(c\right)}{=}\log\left|\mathbf{I}+\left[\begin{array}{c}
\rho^{\left(1-A_{2}\right)}U_{2}\Lambda\\
\rho^{\left(\alpha-A_{2}\right)}U_{1}\Lambda
\end{array}\right]\left[\Lambda U_{2}^{\dagger}\;\Lambda U_{1}^{\dagger}\right]\right|\\
 &  & \negthickspace\negthickspace\negthickspace\overset{\left(d\right)}{=}\log\left|\mathbf{I}_{M_{2}}+\left[\tilde{U}_{2}^{\dagger}\;\tilde{U}_{1}^{\dagger}\right]\left[\begin{array}{c}
\rho^{\left(1-A_{2}\right)}\tilde{U}_{2}\\
\rho^{\left(\alpha-A_{2}\right)}\tilde{U}_{1}
\end{array}\right]\right|\\
 &  & \negthickspace\negthickspace\negthickspace=\log\left|\mathbf{I}_{M_{2}}+\rho^{\left(1-A_{2}\right)}\tilde{U}_{2}^{\dagger}\tilde{U}_{2}+\rho^{\left(\alpha-A_{2}\right)}\tilde{U}_{1}^{\dagger}\tilde{U}_{1}\right|\\
 &  & \negthickspace\negthickspace\negthickspace\overset{\left(e\right)}{=}f\left(M_{2},\left(1-A_{2},N_{2}\right),\left(\alpha-A_{2},N_{1}\right)\right)\log\rho
\end{eqnarray*}
and dividing both sides by $\log\rho$ as $\rho\rightarrow\infty$,
we obtain \eqref{eq:gdof-zic-equation-d2}. In the above derivation,
$\left(a\right)$ is obtained by substituting $Q_{2}=\rho^{-A_{2}}\mathbf{I}_{M_{2}}$,
$\left(b\right)$ uses the SVD of $\left[\begin{array}{c}
H_{22}\\
H_{12}
\end{array}\right]$ defined earlier, $\left(c\right)$ uses the following partitioning,
$U\negmedspace=\negmedspace\left[\begin{array}{c}
U_{2}\\
U_{1}
\end{array}\right]$, where $U_{i}\in\mathbb{C}^{N_{i}\times M_{2}}$, for $i\in\left\{ 1,2\right\} $,
and finally $\left(d\right)$ follows, after defining $\tilde{U}_{i}\triangleq U_{i}\Lambda$,
from the identity $\left|\mathbf{I}+AB\right|=\left|\mathbf{I}+BA\right|$.
The asymptotic approximation in $\left(e\right)$ is a direct consequence
of Lemma \ref{lem:MAC-approx}.

To obtain \eqref{eq:gdof-zic-equation-d2-deta}, we proceed from \eqref{eq:info-theory-rc-ri}
as follows (the conditioning on $\mathcal{H}^{n}$ and the $\mathcal{O}\left(1\right)$
terms are omitted):
\begin{eqnarray*}
 &  & \bar{R}_{c}+\bar{R}_{2}\\
 &  & \leq I\left(X_{c}^{\prime},X_{2}^{\prime};Y_{2}^{\prime}\right)\\
 &  & =h\left(Y_{2}^{\prime}\right)+\mathcal{O}\left(1\right)\\
 &  & =h\left(\sqrt{\rho^{\alpha}}H_{12}X_{2}^{\prime}\right)+h\left(\sqrt{\rho}H_{22}\left(X_{c}^{\prime}\negmedspace+\negmedspace X_{2}^{\prime}\right)\left|\sqrt{\rho^{\alpha}}H_{12}X_{2}^{\prime}\right.\right)\\
 &  & \overset{\left(a\right)}{\geq}\left(\alpha-A_{2}\right)^{+}N_{1}^{\prime}\log\rho\\
 &  & +h\left(\sqrt{\rho}H_{22}\left(X_{c}^{\prime}+X_{2}^{\prime}\right)\left|\sqrt{\rho^{\alpha}}H_{12}X_{2}^{\prime},X_{2}^{\prime}\right.\right)\\
 &  & =\left(\alpha-A_{2}\right)^{+}N_{1}^{\prime}\log\rho+h\left(\sqrt{\rho}H_{22}X_{c}^{\prime}\right)\\
 &  & =\left(\alpha-A_{2}\right)^{+}N_{1}^{\prime}\log\rho+N_{2}\log\rho,
\end{eqnarray*}
where $\left(a\right)$ is true because conditioning reduces entropy,
and any region contained in the achievable region is also achievable.
By dividing both sides above by $\log\rho$ as $\rho\rightarrow\infty$,
we obtain \eqref{eq:gdof-zic-equation-d2-deta}. 

\bibliographystyle{IEEEtran}
\bibliography{bibliography}

\end{document}